\documentclass[]{aa}
\usepackage[varg]{txfonts}

\usepackage[utf8]{inputenc}
\usepackage{graphicx}
\usepackage{amssymb}
\usepackage{gensymb}
\usepackage{textgreek}

\usepackage[version=4]{mhchem}

\usepackage[colorlinks=true, linkcolor=blue, urlcolor=black, citecolor=blue]{hyperref}

\usepackage{natbib}
\bibpunct{(}{)}{;}{a}{}{,}

\renewcommand\makeLineNumber{}

\begin{document}

\title{CO desorption from interstellar icy grains induced by IR excitation of superhydrogenated PAHs}
\titlerunning{CO desorption by IR excitation of superhydrogenated PAHs}

\authorrunning{L. Slumstrup et al.}

\author{ L. Slumstrup \inst{\ref{inst-InterCat}}
    \and J. D. Thrower \inst{\ref{inst-InterCat}}\thanks{Corresponding author: thrower@phys.au.dk}
    \and A. T. Hopkinson \inst{\ref{inst-InterCat}}
    \and G. Wenzel \inst{\ref{inst-Gabi-MIT},\ref{inst-Gabi-Harvard}}
    \and R. Jaganathan \inst{\ref{inst-InterCat}}
    \and J. G. M. Schrauwen \inst{\ref{inst-FELIX}}
    \and B. Redlich \inst{\ref{inst-FELIX}}\thanks{\emph{Present address: Deutsches Elektronen-Synchrotron DESY, 22607 Hamburg, Germany}}
    \and S. Ioppolo \inst{\ref{inst-InterCat}}
    \and L. Hornek{\ae}r \inst{\ref{inst-InterCat}}\thanks{Corresponding author: liv@phys.au.dk} 
    }

\institute{
Center for Interstellar Catalysis, Department of Physics and Astronomy, Aarhus University, 8000 Aarhus C, Denmark \label{inst-InterCat}
\and
Department of Chemistry, Massachusetts Institute of Technology, Cambridge, MA 02139, USA \label{inst-Gabi-MIT}
\and
Center for Astrophysics | Harvard \& Smithsonian, Cambridge, MA 02138, USA \label{inst-Gabi-Harvard}
\and
HFML-FELIX Laboratory, Radboud University, Nijmegen 6525 ED, The Netherlands \label{inst-FELIX}
}

\date{}

\abstract
{Infrared (IR) radiation dominates dense, interstellar clouds, yet its effect on icy grains remains largely unexplored. 
Its potential role in driving the photodesorption of volatile species from such grains has recently been demonstrated, providing a crucial link between the solid state reservoir of molecules and the gas phase. 
}
{In this work, we investigate IR-induced photodesorption of CO for astrophysically relevant ice systems containing perhydropyrene (PHP). This fully superhydrogenated version of pyrene is used as an analogue for large carbonaceous molecules such as polycyclic aromatic hydrocarbons (PAHs) and related species, as well as hydrogenated carbonaceous grains. The abundance and range of strong IR absorption bands of these carbonaceous species make them interesting candidates for IR-induced effects. 
}
{We present IR spectroscopic and mass spectrometric measurements probing the effects of IR radiation on two ice systems: a layered ice with CO on top of PHP, and a CO:PHP mixed ice. These ices were irradiated with IR radiation from the FELIX IR Free Electron Laser (FEL) FEL-2.
} 
{In accordance with previous studies, we confirm that direct excitation of CO is not an efficient pathway to CO desorption, indicating that another energy dissipation mechanism exists. 
We demonstrate that vibrational excitation of the PHP CH stretching modes leads to efficient CO photodesorption. The derived photodesorption yields are an order of magnitude higher for the layered than the mixed system and comparable to those previously obtained for CO photodesorption from CO on amorphous solid water upon excitation of \ce{H2O} vibrational modes.
Our results indicate that IR excitation of carbonaceous molecules and grains in dense clouds could potentially play an important role in the desorption of volatile species such as CO from icy grains. 
}
{}


\maketitle

\section{Introduction}

In dense interstellar clouds, radiation drives physical and chemical processes in molecular ices on dust grains. CO is a key component of these ices, freezing out as a non-polar layer on top of water-rich ice layers in the denser parts of molecular clouds where grain temperatures reach as low as 10\,K. 
Interactions with radiation and atomic species promote chemistry within the icy mantles through the formation of radicals, which can recombine to produce larger molecules, including interstellar complex organic molecules (COMs).
For example, hydrogen addition reactions in the CO ice layer have experimentally been shown to lead to the formation of methanol (\ce{CH3OH}) and species as large as simple sugars and sugar alcohols \citep{Fuchs2009, Fedoseev2017}.  
For more complex, mixed ices, both hydrogen addition and ultraviolet (UV) irradiation have been shown to result in a range of biologically relevant species such as amino acids, sugars, and fatty acids \citep{Ioppolo2021, Oberg2016, MunozCaro2002, Meinert2016, Nuevo2011}.
The observation of gas-phase COMs in low-temperature regions where grain temperatures are too low to lead to thermal desorption of such species indicates the need for additional non-thermal desorption mechanisms, e.g. \citep{Cernicharo2012,Guzman2013,Vastel2014,Perotti2020,Evans2025}. 
Non-thermal desorption has similarly been invoked to explain observations of rotationally cold gas-phase CO in low-temperature regions, e.g. \citep{Willacy2000,Pietu2007,Oberg2015}.
Radiation-induced desorption processes are of great interest in this regard, providing a crucial link between the condensed phase molecular repository found in interstellar ices and the interstellar gas-phase.

UV radiation has been shown to induce desorption of species such as CO, \ce{CH3OH}, \ce{CH4}, and \ce{CH3CN} from the 10\,K icy grain mantles \citep{Oberg2009, Fayolle2011, Bertin2012, Bertin2016, Dupuy2017, Basalgete2021}. However, while UV radiation is abundant in many regions of the interstellar medium (ISM), the UV component of the interstellar radiation field (ISRF) is strongly attenuated within dense, interstellar clouds beyond $A_\mathrm{V}\,{\sim}\,3$ \citep{Mathis1983}. Instead, secondary UV photons from cosmic-ray induced excitation of \ce{H2} molecules are the only source of UV radiation, with an estimated flux on the order of $10^4$\,photons\,cm$^{-2}$\,s$^{-1}$ integrated across the UV spectral range \citep{Cecchi-Pestellini1992,Shen2004}. The infrared (IR) component of the ISRF is less shielded and thus dominates inside dense clouds, with the photon flux estimated to typically be on the order of $10^9$\,\,photons\,cm$^{-2}$\,s$^{-1}$ or higher for all wavelengths in the 1\,{\textmu}m to 10\,{\textmu}m range \citep{Mathis1983,Porter2005,Roueff2013}. 
While UV irradiation of icy grain analogues has been investigated extensively, the effects of IR radiation on such ices have received less attention. Beyond its role in radiative transport, the influence of IR radiation on condensed-phase chemistry and physical effects such as ice restructuring and desorption remains poorly understood. The abundance of IR radiation, combined with the strong IR absorbance of common interstellar ice species such as \ce{H2O} and \ce{CO}, makes these processes highly relevant for astrochemistry studies.

IR irradiation has been shown to promote ice restructuring, leading to a more ordered structure, e.g. for amorphous solid water (ASW) \citep{Noble2014a, Noble2014b, Noble2020_a, Cuppen2022_ASW, Coussan2015, Coussan2022} and \ce{CO2} \citep{Ioppolo2022}. IR-induced desorption has been found to occur for crystalline \ce{H2O} ice \citep{Noble2020_a,Krasnopoler1998,Focsa2003} and CO from CO-\ce{H2O} mixtures following excitation of \ce{H2O} vibrational modes \citep{Ingman2023}. 
Recently, we have used the FELIX IR Free Electron Laser (FEL) FEL-2 to investigate the IR-induced CO photodesorption from CO on ASW \citep{Slumstrup2025}, simulating the simple CO and \ce{H2O} layered ices in interstellar clouds.
By employing a combination of IR spectroscopy and mass spectrometric detection of desorbing molecules, we demonstrated that excitation of the stretching mode of CO leads only to inefficient CO desorption for both pure CO ice and CO on ASW. In contrast, excitation of vibrational modes of the underlying ASW resulted in efficient CO desorption and even some desorption of the \ce{H2O} itself. 
These results suggest that indirect non-thermal desorption, where energy absorbed by one species triggers the desorption of another, may play a significant role in the release of volatile species such as \ce{CO}.

Another important class of interstellar molecules that absorb strongly in the IR is the family of polycyclic aromatic hydrocarbons (PAHs). Indeed, they are abundantly detected in the gas-phase through their strong mid-IR emission following UV excitation with characteristic features at 3.3, 6.2, 7.7, 8.6, 11.2, 12.7, and 16.4\,{\textmu}m \citep{Tielens2008}.
PAHs can be considered as an extension of the size distribution of interstellar carbonaceous dust grains to the molecular scale and are estimated to account for 10-20\% of the galactic carbon budget \citep{JoblinTielens2011}. As such, they are suggested to play an important role in the physics and chemistry of the ISM. PAHs have e.g. been demonstrated to act as catalysts in the formation of \ce{H2} \citep{RaulsHornekaer2008,Thrower2012,Mennella2012,Cazaux2016,Jensen2019} in radiation-dominated regions of the ISM where dust grain formation routes are inefficient to explain the observed \ce{H2} abundances \citep{Habart2003}. The formation of \ce{H2} takes place through the addition to and abstraction of H atoms from the PAHs, leading to the formation of superhydrogenated PAHs (HPAHs) which are also detected in the ISM through e.g. their 3.4 and 6.9\,{\textmu}m emission features \citep{Sloan1997,Steglich2013,Sandford2013}. 
The range of strong vibrational modes of PAHs and their derivatives makes them well-suited for studying their interaction with IR radiation. In interstellar clouds, PAHs are expected to freeze out onto the dust grains \citep{Chiar2021}, either becoming part of the grains themselves before ice formation or becoming embedded in the icy mantles. 
Based on observations, it is estimated that interstellar ices could contain PAH-related molecules at the 2-3\% level relative to water \citep{Bouwman2011_I}. Here, PAHs contribute to the solid-state chemical complexity through interactions with UV radiation and protons, leading to HPAHs and the addition of other functional groups \citep{Bouwman2011_I,Bernstein1999,Bernstein2003,Gudipati2012,Noble2020_b}. 
In these studies, PAHs have mainly been considered as interacting with \ce{H2O}. However, PAHs have limited collision rates with the icy grains because of their low concentration and mobility in the gas phase and are therefore unlikely to fully freeze out in the \ce{H2O} rich ice layer, according to a model by \citet{Bouwman2011_II}. For that reason, PAHs can be expected to also freeze out later in the cloud lifetime, where the CO-rich ice layer forms on the grains. It is therefore relevant to consider whether IR absorption by PAH molecules in close proximity to CO molecules could drive CO desorption.

In this work, we extend our study of IR-induced photodesorption of CO to mixed and layered ices containing CO and PAHs, using a superhydrogenated form of pyrene (\ce{C16H10}). Pyrene is known to exist in the ISM as it has recently been detected in a dense cloud through its cyano-derivatives \citep{Wenzel2024,Wenzel2025a}. Pyrene is also found in solar system bodies such as meteorites \citep{Sephton2002} and asteroids \citep{Naraoka2023,Zeichner2023} which are considered remnants of the early solar system.
Pyrene is well-studied experimentally, e.g. embedded in \ce{H2O} ices \citep{Hardegree-Ullman2014, Chiar2021} where photo-processing was found to lead to superhydrogenated and ionised pyrenes and oxygen-containing derivatives \citep{Bouwman2011_I,Cuylle2014,Cook2015}.
The fully superhydrogenated pyrene, perhydropyrene (PHP; \ce{C16H26}), was chosen as an analogue in this work because of its strong aliphatic C-H stretching vibrational modes at 3.42\,{\textmu}m which allow for strong absorption at high photon energies. These vibrational modes also have the advantage that they do not overlap with any modes of \ce{H2O}, and hence related studies with \ce{H2O}-containing ices can examine the excitation of PHP without also exciting any \ce{H2O} modes. 
In particular, we examine the excitation of both PHP and CO in two systems, a CO:PHP mixed ice and a CO on PHP layered ice. In the mixed system, PHP acts as an analogue for large cyclic hydrocarbon molecules such as (H)PAHs embedded in CO ice. 
In the layered ice system, PHP acts as an analogue for hydrogenated carbonaceous dust grains.
We show that efficient CO photodesorption occurs following excitation of PHP in both systems and derive photodesorption cross-sections and yields. We discuss the astrophysical implications of these results, which indicate a potential role of IR radiation in determining the ice-gas balance of dense clouds and in governing the complex chemistry occurring within.

\section{Experimental details}

The experiments described in this work were performed at the Laboratory Ice Surface Astrophysics (LISA) Ultra-High Vacuum (UHV) end-station at the HFML-FELIX facility in Nijmegen, the Netherlands, during a beamshift in June 2024. The setup is described in detail elsewhere \citep{Ioppolo2022,Schrauwen2024}. 
Briefly, the LISA end-station consists of a UHV chamber with a base pressure of approximately $5\times 10^{-10}$\,mbar during cryostat operation with the sample at its base temperature. Ices were grown on a gold-plated copper substrate attached to the cryostat. The temperature of the substrate was 9\,K throughout ice formation and during experiments as measured with a silicon diode attached to the base of the substrate. PHP, formally hexadecahydropyrene, (\ce{C16H26}) (TCI Europe; purity >\,97\%) was introduced to the chamber via a fine leak valve and home-built stainless steel dosing tube directed at the sample with a tube-to-sample distance of $\sim$\,5\,cm. PHP is a low vapour pressure liquid at room temperature and was purified through pumping cycles with a dedicated high-vacuum pumping station to remove air and water. CO (Linde HiQ grade; purity 99.997\%) was deposited through background deposition. 
The CO:PHP mixed ice was grown through co-deposition of CO and PHP for 2000\,s to ensure effective mixing of the species, with a total chamber pressure of $6\times 10^{-8}$\,mbar, where PHP and CO accounted for approximately $2\times 10^{-10}$ and $5\times 10^{-8}$\,mbar, respectively. 
For the layered ice, the deposition parameters for the two species were chosen such that the amounts deposited were similar to the mixed ice. PHP was dosed for 1000\,s at double the rate used for the mixed ice, as determined through its partial pressure in the chamber, monitored by a quadrupole mass spectrometer (QMS; Hiden Analytical Hiden HAL/3F RC PIC). Subsequently, CO was dosed on top of the PHP for 1400\,s at a total chamber pressure of $7\times 10^{-8}$\,mbar.
These deposition procedures resulted in around 19 layers of CO and a CO-to-PHP ratio of ${\sim}10$ on the surface, as detailed in the following. Under these deposition conditions, CO is deposited as an amorphous ice \citep{Kouchi2021}, and we expect the same for the PHP. 

Reflection-Absorption IR Spectroscopy (RAIRS) was used to monitor ice growth and the effects of FEL irradiation. The RAIRS setup consists of a Bruker Vertex 80v Fourier Transform IR (FTIR) spectrometer with the IR beam interfaced to the UHV chamber. 
The IR beam is incident at a grazing angle of $10\degree$ to the substrate surface. An aperture of 1\,mm was used, resulting in a RAIRS spot size on the surface of 0.92\,cm$^2$ as detailed in Appendix \ref{App:SpotAreas}. RAIR spectra were recorded in the 5000-500\,cm$^{-1}$ range with a resolution of 0.5\,cm$^{-1}$, and averaged through the co-addition of 128 or 256 scans. 
All spectra were baselined with a global cubic spline prior to plotting and analysis, and local straight baselines are used for integration of peaks.

Following deposition, the ices were irradiated using the free electron laser FEL-2, which is tunable in the wavelength range of 2.7-45\,{\textmu}m. 
The FEL-2 beam is monochromatic with a spectral FWHM estimated to be $\sim 0.8\%\ \delta \lambda/\lambda$. For our experiments, the beam consisted of a series of 6\,{\textmu}s long macropulses at a frequency of 5\,Hz. Each macropulse comprised a train of micropulses with a repetition rate of 1\,GHz. The micropulses were therefore spaced 1\,ns apart. 
The \textit{p}-polarized component of the FEL beam was incident on the substrate at an angle of $45\degree$ to the surface. The irradiation energy, $E_\mathrm{irr}$, i.e. the total energy of each macropulse, varied between irradiations in the range of 35-61\,mJ.
Ices were irradiated with the FEL-2 tuned to 3.42\,{\textmu}m and 4.67\,{\textmu}m, exciting the PHP CH stretching mode and CO stretching modes, respectively. Additionally, an off-resonance reference irradiation was performed at 4.02\,{\textmu}m for the mixed ice. 
The substrate can be moved vertically such that a series of pristine sample spots are available on the surface for irradiation. Each individual irradiation measurement was performed on a previously un-irradiated spot for a duration of 1\,min. 
Subsequent irradiation of previously irradiated spots confirmed that no further changes were observed in the RAIR spectra.

The fluence of incident photons in each macropulse of FEL-2 is determined as
\begin{equation} \label{eq:phi}
    \phi = \frac{E_\mathrm{irr}}{E_\mathrm{photon}} \frac{r_\mathrm{macro}}{S_\mathrm{FEL}},
\end{equation}
where $E_\mathrm{photon} = hc/\lambda$ is the photon energy at wavelength $\lambda$. The first fraction then corresponds to the number of photons in each macropulse. $r_\mathrm{macro}=5$ Hz is the macropulse frequency and $S_\mathrm{FEL}$ is the area of the irradiated spot on the substrate. This area varies with irradiation wavelength, ranging from 0.0077 to 0.0143\,cm$^2$ for the irradiation wavelengths used in this work. 
In all cases, the irradiation spots are significantly smaller than the RAIRS spot, meaning that only a fraction of the ice observed with RAIRS was exposed to the FEL irradiation. Details are provided in Appendix \ref{App:SpotAreas}. 

In addition to recording RAIR spectra prior to and after FEL irradiation, the QMS was used to directly detect species desorbing from the surface during irradiation. The QMS effective time resolution was optimised to 10\,ms by restricting the mass filter to detect only CO ($m/z = 28$). The QMS has a processing period on the order of a few 10ths of a ms associated with each measurement point during which it is not acquiring data, resulting in an incomplete recording of the full desorption event.

\section{Results}

\subsection{RAIR spectra and composition of the as-deposited ices}
\begin{figure*}[t]
    \centering
    \includegraphics{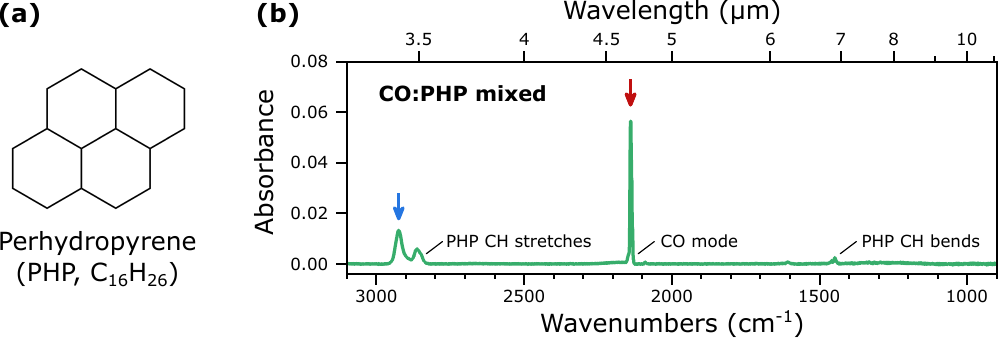}
    \caption{(a) The structure of perhydropyrene (PHP, \ce{C16H26}), the fully superhydrogenated version of pyrene.
    (b) RAIR spectrum of the CO:PHP mixed ice. 
    The main spectral features are marked, and coloured arrows indicate the irradiation wavelengths: the strongest CH stretch of PHP at 3.42\,{\textmu}m (blue) and the CO stretching mode at 4.67\,{\textmu}m (red). 
    }
    \label{fig:IRSpectrum}
\end{figure*}

PHP, \ce{C16H26} -- the fully aliphatic form of pyrene -- was used as an analogue for large cyclic hydrocarbon molecules such as PAHs as well as hydrogenated carbonaceous grains. Its molecular structure is shown in Fig. \ref{fig:IRSpectrum}(a). Figure \ref{fig:IRSpectrum}(b) depicts the RAIR spectrum of a mixed CO:PHP ice, deposited at 9\,K on an Au-plated copper substrate. The spectrum of the layered ice is omitted for clarity, as the PHP bands exhibit no significant differences compared to the mixed ice. 
To the best of our knowledge, this is the first experimental IR spectrum of PHP obtained in the condensed phase. The spectrum is dominated by the aliphatic CH stretching modes at 3.4 and 3.5\,{\textmu}m with weaker bending modes at 6.9\,{\textmu}m. These aliphatic bands are shifted from the corresponding aromatic bands of e.g. pyrene and its partially hydrogenated derivatives such as hexahydropyrene (HHP, \ce{C16H16}) at 3.3 and 7.0\,{\textmu}m \citep{Chiar2021}. 
The wavelengths of the PHP bands are consistent with the aliphatic bands of HHP \citep{Chiar2021} and the general characteristic aliphatic modes of superhydrogenated PAHs \citep{Sloan1997,Steglich2013,Sandford2013}.
There is no overlap between any bands of the CO and PHP across the mid-IR range. Coloured arrows in Fig.~\ref{fig:IRSpectrum} indicate the wavelengths used for FEL irradiations of the ices. These correspond to the strongest aliphatic CH stretch of PHP at 3.42\,{\textmu}m (2920\,cm$^{-1}$; blue arrow) and the CO stretching mode at 4.67\,{\textmu}m (2140\,cm$^{-1}$; red arrow). The mixed ice was also irradiated at an off-resonance reference wavelength of 4.02\,{\textmu}m (2490\,cm$^{-1}$). We use this colour scheme throughout the paper. The irradiation wavelengths are summarised in Table \ref{tab:overview}.

\begin{table*}[ht]
    \caption{Overview of the experiments performed in this work with column densities, $N$, and irradiation wavelengths, $\lambda_\mathrm{FEL}$, for each ice.}
    \label{tab:overview}
    \centering
    \begin{tabular}{l c c c}
    \hline \hline
    \begin{tabular}[t]{c} Ice \end{tabular} &
    \begin{tabular}[t]{c} $N_\mathrm{CO}$ \\ ($10^{15}$\,molecules\,cm$^{-2}$)\end{tabular} &
    \begin{tabular}[t]{c} $N_\mathrm{PHP}$\\ ($10^{15}$\,molecules\,cm$^{-2}$)\end{tabular} &
    \begin{tabular}[t]{c} Irradiation wavelength \\ $\lambda_\mathrm{FEL}$ ({\textmu}m) \end{tabular}
    \\
    \hline
    CO:PHP mixed        &  $19.1\pm 0.2$        &  $ 2.0 \pm 0.4  $ &  4.02, 3.42, 4.67  \\
    CO/PHP layered      &  $ 18.47 \pm 0.01$    &  $ 1.7 \pm 0.1  $ &  3.42, 4.67     \\
    \hline
    \end{tabular}
    \tablefoot{The layered system has CO on top of PHP. Column density values are averaged over the irradiation spots on the sample with the errors reflecting the standard deviation. No band strength uncertainty has been taken into account.}
\end{table*}

The amount of each ice component is determined through the column density, $N$, found from the integrated absorbance, $\int \mathrm{Abs}$, of an IR band with respect to wavenumber, $\tilde{\nu}$, and the transmission band strength, 
$A_\mathrm{band}$, using
\begin{equation} \label{eq:ColDens}
    N = \ln{10} \ \frac{\int \mathrm{Abs}(\tilde{\nu}) \ \mathrm{d}\tilde{\nu}}{A_\mathrm{band} \times 4.5}.
\end{equation}
The sensitivity enhancement from RAIRS relative to transmission IR is taken into account by the factor of 4.5, computed for CO and ASW ices at the LISA setup. The details of this can be found in our previous work \citep{Slumstrup2025}. 
For the CO band at 4.67\,{\textmu}m, a transmission IR band strength of $A_\mathrm{band}=1.12\times 10^{-17}$\,cm/molecule for pure CO is used \citep{Bouilloud2015}. 
We estimate a band strength of $A_\mathrm{band}=1.53\times 10^{-16}$\,cm/molecule for the combined CH stretching bands of PHP based on comparison with RAIRS and QMS measurements of pyrene at the LISA setup, as there are no published values for PHP. See Appendix \ref{App:PHPamounts} for details. 

Table \ref{tab:overview} reports the derived column densities for the two CO-PHP ice systems. 
The column densities of CO and PHP are similar for the two ices, with a CO-to-PHP ratio of ${\sim}\,10$. 
Assuming a typical molecular layer density for small molecules of $10^{15}$\,cm$^{-2}$, both ices contain around 19 layers of CO. For the larger PHP molecule, we approximate a layer density of $0.26\times 10^{15}$\,cm$^{-2}$ based on its molecular geometry as discussed in Appendix \ref{App:PHPamounts}. Accordingly, we determine that both ice systems contain around seven layers of PHP. For the layered system, this means that no CO molecules are expected to be in contact with the Au-plated copper substrate.

\subsection{QMS detection of IR-induced CO desorption}

\begin{figure}[t]
    \centering
    \includegraphics{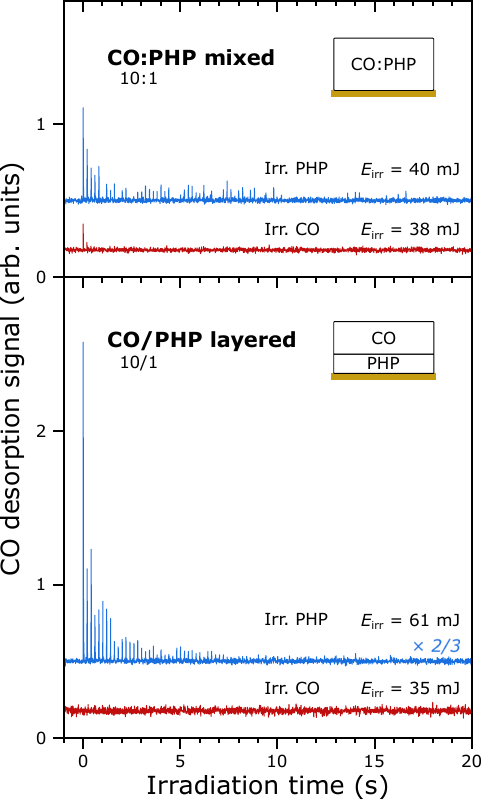}
    \caption{CO desorption signals detected by the QMS for $m/z=28$ during irradiations of the mixed CO:PHP and layered CO/PHP ices, baselined and offset for clarity.
    Irradiation energies, $E_\mathrm{irr}$, are denoted above each spectrum.
    For better visual comparison, the desorption trace for the PHP mode irradiation of the layered ice with higher irradiation energy has been scaled by a factor of 2/3. 
    Each individual desorption peak corresponds to a 5\,Hz macropulse of the FEL. The zero-time for irradiation is defined as the onset of the first CO desorption peak.}
    \label{fig:QMSmain}
\end{figure}

Figure~\ref{fig:QMSmain} shows the CO desorption detected with the QMS during FEL irradiation exciting the CO and PHP vibrational modes for the mixed and layered ices. 
Desorption traces have been baselined to the average level of CO signal prior to irradiation and offset from each other for clarity. Each individual desorption peak corresponds to a 5\,Hz macropulse of the FEL. All irradiations were continued for 1\,min, although most desorption occurred within the initial ${\sim}\,10$\,s seconds, as evident from Fig.~\ref{fig:QMSmain}.
All irradiations were performed on spots of the ices with similar CO-to-PHP ratios of ${\sim}\,10$. Irradiation energies $E_\mathrm{irr}$ are denoted above each trace with variations arising due to the varying performance of the FEL during the beamshift.
To facilitate comparison of the desorption data for the two ices, the desorption trace for exciting the PHP mode for the layered ice in Fig.~\ref{fig:QMSmain} has been scaled by a factor of 2/3, to account for the higher irradiation energy. 
This irradiation still shows the most significant CO desorption signal. The remaining figures and analysis of this irradiation use the non-scaled data.

As expected, the QMS signal shows no CO desorption from the ices before the start of irradiation. 
Similarly, the 4.02\,{\textmu}m off-resonance irradiation for the mixed ice resulted in no CO desorption, as shown in Appendix \ref{App:QMSpeaks}.
Irradiating at 4.67\,{\textmu}m, corresponding to the CO mode, (red traces in Fig.~\ref{fig:QMSmain}) results in a weak desorption signal from the mixed ice. No CO desorption was observed for excitation of the CO mode for the layered ice. This indicates that direct excitation of CO molecules is not efficient in driving CO desorption. There must therefore exist alternative, non-desorptive pathways for the dissipation of energy absorbed directly by CO molecules.
For both the mixed and layered ices, irradiating at 3.42\,{\textmu}m, corresponding to the PHP mode, (blue traces in Fig.~\ref{fig:QMSmain}) leads to significant CO desorption, with the strongest desorption for the layered ice. This indicates that energy transfer occurs between the two species despite them having no spectral overlap of vibrational modes in the mid-IR range, as illustrated in Fig. \ref{fig:IRSpectrum}. 
In the case of the mixed CO:PHP ice, a fraction of CO molecules is likely trapped beneath PHP molecules, reducing the desorption efficiency, consistent with the difference in CO desorption seen in Fig.~\ref{fig:QMSmain}. 
For the layered ice, most of the CO molecules are not in direct contact with a PHP molecule. The strong desorption signal therefore indicates that energy can transfer through the CO layers to the vacuum interface.
The individual desorption peaks are shown in detail in Appendix \ref{App:QMSpeaks} along with a description of the integration process used in the following.

\begin{figure}[t]
    \centering
    \includegraphics{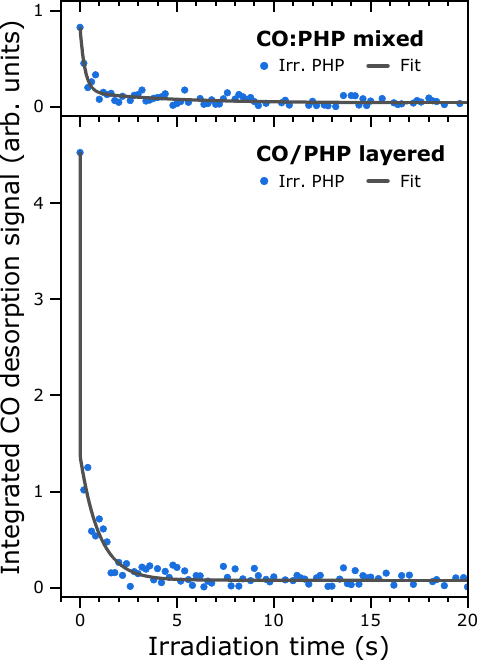}
    \caption{Bi-exponential fits to the integrated CO desorption signals derived from the data presented in Fig. \ref{fig:QMSmain}, for the 3.42\,{\textmu}m irradiation of the mixed (top panel) and layered (bottom panel) CO-PHP systems. The data for the layered ice has not been scaled.}
    \label{fig:QMSfit}
\end{figure}

To determine the desorption rate for irradiation at 3.42\,{\textmu}m, resonant with the PHP mode, we integrate the CO desorption signal for each macropulse, for both the mixed and layered ices shown in Fig.~\ref{fig:QMSmain}. 
The integration process is described in Appendix \ref{App:QMSpeaks}. The resulting integrated CO desorption signal is then plotted as a function of irradiation time in Fig.~\ref{fig:QMSfit}. In this case, no scaling has been applied to the desorption data for the layered ice. 
The desorption signals appear to decay exponentially with time, indicating a first-order desorption process with respect to $N_\mathrm{CO}$. This behaviour has previously been observed for IR-induced CO desorption from CO on \ce{H2O} \citep{Slumstrup2025} and CO-\ce{H2O} mixtures \citep{Ingman2023} following excitation of various \ce{H2O} vibrational modes. First-order desorption kinetics are typically associated with desorption from a monolayer of adsorbate, where the desorption rate depends on the remaining coverage. The apparent first-order desorption kinetics for our thicker ice layers is therefore surprising. For the mixed ice, this might indicate that a limited population of the adsorbed CO ice can desorb. For the layered ice, the effective CO ice thickness will depend on the roughness of the deposited PHP layer. A rough surface would have a larger surface area and, therefore, the CO would be more thinly spread over the surface, potentially resulting in first-order desorption kinetics.
Further measurements probing the dependence of the CO desorption efficiency on the thickness of both ice layers and mixing ratio are clearly necessary. Nevertheless, we can use first-order desorption kinetics to determine the desorption efficiency for the specific ices considered in this study. \citet{Ingman2023} found that two decay components were necessary to fit their experimental data, attributing the presence of fast and slow decay components to different desorption processes.
We have followed a similar approach in our analysis of the present data. 
The resulting decays are fitted with a bi-exponential decay function of the form
\begin{equation} \label{eq:bi-exp}
    I(t) = I_1 \exp \left( \frac{-t}{\tau_1} \right) + I_2 \exp \left(\frac{-t}{\tau_2} \right) + I_\infty,
\end{equation}
where $I(t)$ is the integrated desorption signal at time $t$, $I_1$, $I_2$ are the initial signals at $t=0$ and $\tau_1$ and $\tau_2$ are the time constants for the fast and slow components, respectively. The residual baseline QMS signal intensity, $I_\infty$, is included to account for a slight non-zero background level at long times. 
The time constants for the fast and slow desorption processes are given by $\tau_i = 1/ \left( \phi \sigma_i \right)$ where $\sigma_i$ is the desorption cross-section in cm$^2$ and $\phi$ is the photon flux in cm$^{-2}$\,s$^{-1}$.
Per convention, the decay is expressed in terms of time rather than pulse count, but since the experimental photon fluxes also take the pulsed nature of the irradiation into account, this does not change the resulting cross-sections.
The resulting fits are shown as solid lines in Fig.~\ref{fig:QMSfit}.

\begin{table*}[t]
\caption{CO desorption time constants, $\tau$, and photodesorption cross-sections for the slow desorption process, $\sigma_2$.}
\label{tab:QMSfit}
\centering
\begin{tabular}{l c c c c c c}
\hline \hline
    Ice &
    \begin{tabular}[t]{c} $E_\mathrm{irr}$ \\ (mJ) \end{tabular} &
    \begin{tabular}[t]{c} $\phi$ \\  (photons\,cm$^{-2}$\,s$^{-1}$) \end{tabular} &
    \begin{tabular}[t]{c} $\tau_1$ \\ (s) \end{tabular} &
    \begin{tabular}[t]{c} $\tau_2$ \\ (s) \end{tabular} &
    \begin{tabular}[t]{c} $\sigma_2$ \\ (cm$^2$) \end{tabular} \\
\hline
CO:PHP mixed &   40 &   $4.5\times 10^{20}$ &      $0.25\pm 0.04$ &             $4.4\pm1.3$ &   $(0.5 \pm 0.2) \times 10^{-21}$ \\
CO/PHP layered &   61 &   $6.9\times 10^{20}$ &    $0.001$\tablefootmark{a} &   $1.0\pm0.1$ &   $(1.5 \pm 0.1)\times 10^{-21}$ \\ 
\hline
\end{tabular}
\tablefoot{Values are obtained from the bi-exponential fits in Fig.~\ref{fig:QMSfit} for irradiation at 3.42\,{\textmu}m, corresponding to the PHP band. The associated irradiation energies, $E_\mathrm{irr}$, and photon fluxes, $\phi$, are also listed. Error limits represent one standard deviation on the fitted value.\\
\tablefoottext{a} {The fit is insensitive to this time constant due to its small value and its uncertainty is therefore not reported.} 
}
\end{table*}

Table \ref{tab:QMSfit} shows the resulting time constants $\tau_i$ from the fits, along with the relevant photon fluxes $\phi$. In addition, the corresponding photodesorption cross-sections for the slow process, $\sigma_2$, for each irradiation, are given. 
For both ices, $\tau_2$ is significantly larger than $\tau_1$, with the fast process occurring on a sub-second time scale while the slow process occurs over several seconds.
For both processes, the desorption from the layered ice is considerably more efficient than from the mixed.
Although the two-component exponential decay function in general fits the experimental data well, for the layered ice the fit is insensitive to the exact value of $\tau_1$. We attribute this to the actual time constant being of the same order as the time between successive macropulses and thus consider the value quoted an upper limit.
Because of this, no reliable desorption cross-sections for the fast process can be reported. 
For both ices it is clear that there is some residual desorption intensity, $I_\infty$, as some weak desorption peaks are seen throughout the irradiation period. This could be attributed to a combination of the continued slow desorption process, or e.g. from a small thermal desorption component. Continued CO desorption could also occur from a continuous re-filling of desorption sites either from diffusion into the irradiated area or a small amount of re-deposition from the gas phase. For both irradiations resonant with the PHP mode, we performed a second equivalent irradiation on the same spot to check for any long-term effects. In both cases, only a few desorption peaks at the level of $I_\infty$ were observed. 
Our results indicate that significantly more CO desorbs, per photon, following the excitation of the PHP mode in the layered compared to the mixed ices, despite the similar initial amounts of CO.
A comparison of the cross-sections for the slow process, specifically, shows that the probability for photodesorption of CO following a PHP excitation is around 3 times higher for the layered than for the mixed ice.

\subsection{Quantification of CO desorption through RAIRS}

\begin{figure*}
    \centering
    \includegraphics{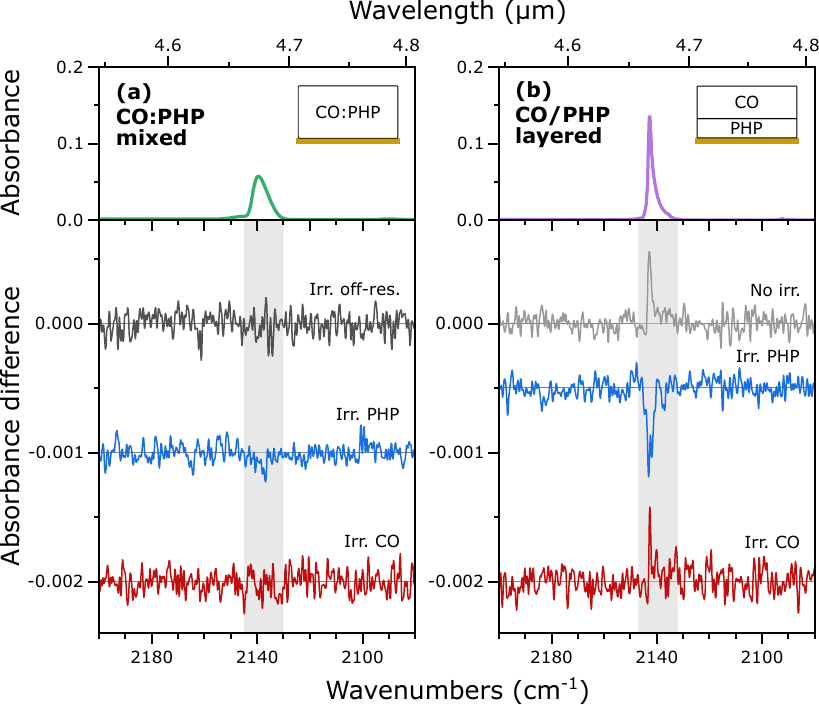}
    \caption{
    RAIR spectra (top) and difference spectra (bottom) for irradiation of the (a) CO:PHP mixed ice and (b) CO/PHP layered ice, focusing on the region around the CO stretching mode. 
    The CO peak shape is different for the mixed and layered ices but represents the same total area and CO column density.
    The difference spectra represent the irradiations for which QMS data are shown in Fig.~\ref{fig:QMSmain} and Appendix Fig.~\ref{AppFig:QMSzoom}, with the band irradiated denoted above each spectrum. 
    The difference spectrum for a wait time of 1\,min without irradiation for the layered ice (grey) indicates continuous CO deposition. 
    Each difference spectrum has been baseline corrected by subtracting the mean value in the displayed range, excluding the shaded region around the CO band, and has been offset for clarity, represented by horizontal lines.
    The off-resonance irradiation energy was $E_\mathrm{irr}=56\,$mJ.}
    \label{fig:IRdiff}
\end{figure*}

The total amount of CO desorbed during any given irradiation can be determined by considering the IR spectral changes.
Figure \ref{fig:IRdiff} shows the RAIR spectra around the CO stretching band at 2140\,cm$^{-1}$ for (a) the CO:PHP mixed and (b) the CO/PHP layered ice together with the difference spectra corresponding to irradiation at (3.42\,{\textmu}m) and (4.67\,{\textmu}m), exciting the PHP and CO modes, respectively. In addition, difference spectra for the off-resonance irradiation (4.02\,{\textmu}m) of the mixed ice and an equivalent waiting period without irradiation for the layered ice are also shown. 
The shape of the CO band is dependent on the local environment of the CO molecules. For the layered ice, the band closely resembles that of pure CO \citep{Slumstrup2025}. This confirms that most of the CO molecules in the layer are not in direct contact with PHP, indicating that the PHP layer has a low porosity. In the case of the mixed ice, the band is broadened through interactions with PHP. The integrated CO band area is, nevertheless, the same for both ices, representing a CO column density of $\sim 19\times 10^{-15}$\,cm$^{-2}$ as detailed in Table \ref{tab:overview}. 

The difference spectra represent the irradiations for which QMS data are shown in Figs. \ref{fig:QMSmain} and \ref{AppFig:QMSzoom}.
They are obtained by subtracting the IR spectra before irradiation (or wait) from those obtained after. The difference spectra have been baseline corrected by subtracting the mean value in the displayed range, excluding the shaded region around the CO band, and are offset for clarity.
For the CO/PHP layered ice, the difference spectrum in Fig. \ref{fig:IRdiff}(b) obtained after a wait time of 1\,min without irradiation shows a positive feature indicating a continuous CO deposition during the measurements. This deposition resulted from a higher background level of CO in the chamber during the measurements for the layered ice.
The difference spectrum for the wait was obtained immediately before that corresponding to irradiation of PHP, for the same time period, and we therefore assume a similar amount of deposition occurred during the irradiation, where the FEL is only active for 0.003\% of the irradiation time and so should not significantly affect the deposition. 
The difference spectra for the irradiation of the layered ice therefore represent a combination of irradiation-induced changes and this continued deposition. This is evident in the difference spectrum for irradiation of the CO band for the layered ice which strongly resembles that associated with a wait time of 1\,min without irradiation, despite being obtained 30\,min later. This is consistent with the QMS data in Fig.~\ref{fig:QMSmain} that shows this irradiation resulted in no CO desorption.
From Fig. \ref{fig:IRdiff}(b), the deposition appears to be of a similar magnitude as the desorption for the PHP irradiation, which could affect the desorption by continuously supplying new weakly bound CO molecules in the irradiation spot. It should be noted, however, that only a fraction of the deposition observed here for the full RAIRS spot takes place within the much smaller FEL spot and so we do not expect the deposition to affect the desorption process. This is discussed further in the following. 

For both ices, excitation of the PHP mode (3.42\,{\textmu}m) results in CO desorption as revealed by the presence of negative CO stretching band features in the difference spectra in Fig. \ref{fig:IRdiff}. This desorption signature is significantly stronger for the layered ice, consistent with the CO desorption signals measured by the QMS data for the same irradiations shown in Fig. \ref{fig:QMSmain}.
Although the QMS data showed a small CO desorption signal for excitation of the CO mode for the mixed ice, no clear desorption feature can be seen in the IR difference spectrum. This is consistent with an inefficient desorption process.

We note that the IR difference spectra showed small decreases in the intensity of PHP bands following irradiations that caused CO desorption. However, it is unclear if this indicates PHP desorption, as the spectral changes were too weak to be characterised and could feasibly also result from changes in environment following CO desorption. No PHP desorption was detected with the QMS when tuned to its corresponding mass-to-charge ratio ($m/z=218$). The QMS has a lower sensitivity for heavy species, likely making detection of a small amount of desorbing PHP challenging.

\subsection{CO desorption efficiency} 

In the following, we quantify the amount of CO desorbed during irradiation, utilising the data analysis procedures developed in our previous work for the \ce{CO}-\ce{H2O} layered ice system \citep{Slumstrup2025}. The RAIR difference spectra in Fig. \ref{fig:IRdiff} are integrated within the CO peak range represented by the shaded areas. This range is $2145$\,-$2130$\,cm$^{-1}$ for the mixed ice and $2147$\,-$2132$\,cm$^{-1}$ for the layered ice, in order to account for the differences in peak shape for the two systems.
The change in column density, $\Delta N$, can then be found as
\begin{equation} \label{eq:coldens_diff}
    \Delta N = -\ln{10} \ \frac{\int \Delta\mathrm{Abs}(\tilde{\nu}) \mathrm{d}\tilde{\nu} }{A_\mathrm{band} \times 4.5} \frac{S_\mathrm{RAIRS}}{S_\mathrm{FEL}}.
\end{equation}
This relation is equivalent to eq. \eqref{eq:ColDens} with $\Delta\mathrm{Abs}(\tilde{\nu})$ corresponding to the IR band in the difference spectrum.
The additional term, $S_\mathrm{RAIRS}/S_\mathrm{FEL}$, accounts for the RAIRS and FEL beams having different spot sizes on the substrate, resulting in only a few percent of the ice probed by RAIRS being exposed to the FEL beam. The IR difference spectrum therefore averages any changes as if they had occurred over the entire RAIRS spot. This can be corrected by scaling the amount of CO desorbed by the ratio of the RAIRS spot size to that of the FEL. We refer to this as the dilution factor. For the $3.42$\,\textmu m PHP irradiations, the dilution factor is 120, while for $4.67$\,\textmu m CO irradiations it is 64. The difference arises due to the wavelength-dependent size of the FEL beam. See Appendix \ref{App:SpotAreas} for details.

\begin{table*}[t]
\centering
\caption{Desorption amounts and yields for the PHP irradiations of the mixed and layered ices. 
}
\label{tab:results}
\begin{tabular}{l c c c c c c}
\hline \hline
    Ice &
    \begin{tabular}[t]{c} Decrease in \\ column density \\ $\Delta N_\mathrm{CO}$ \\  ($10^{15}$\,mol.\,cm$^{-2}$) \end{tabular} &
    \begin{tabular}[t]{c} Effective \\ irradiation \\ time \\ (s) \end{tabular} &
    \begin{tabular}[t]{c} Desorption yield \\ $Y_\mathrm{des,inc}$ \\ ($10^{-6}$\,mol./ \\ inc. photon) \end{tabular} &
    \begin{tabular}[t]{c} Quantum yield \\ $Y_\mathrm{des,abs}$ \\ ($10^{-4}$\,mol./ \\ abs. photon) \end{tabular} & \\
\hline
CO:PHP mixed &       $5\pm 2$     & 13 &  $0.8\pm 0.4$ &   $0.3 \pm 0.1$ \\
CO/PHP layered &     $16 \pm 2$   &    3 &  $7.7\pm 1.2$ &   $3.1 \pm 0.5$ \\
\hline
\end{tabular}
\tablefoot{The effective irradiation times are derived from the fits to the QMS data. Desorption yield errors are derived from the estimated errors on the decrease in column density.}
\end{table*}

Based on this analysis, we estimate that $\Delta N = (5\pm2)\times 10^{15}$\,molecules\,cm$^{-2}$ and $(16\pm2)\times 10^{15}$\,molecules\,cm$^{-2}$ of CO desorbed following the PHP irradiations of the mixed and layered ice, respectively. For the layered ice, the value has been corrected for the increase in column density during the wait, assuming the deposition continues at a similar rate throughout the irradiations. Considering the dilution factor, the amount of CO deposited within the FEL spot corresponds to only $0.003\%$ of the amount of CO desorbed, i.e. we can assume that the continuous deposition does not affect the desorption process. 
The off-resonance irradiation of the mixed ice resulted in no CO desorption, although integrating the associated RAIR difference spectrum yields a non-zero value of $\Delta N=2\times 10^{15}$\,molecules\, cm$^{-2}$ most likely resulting from the poor signal-to-noise ratio. We therefore take this value as a measure of the noise variation and use it as the uncertainty for the observed CO desorption amounts. Small uncertainties in the RAIRS and FEL beam sizes are accounted for by this conservative error estimate. For example, considering a variation in the RAIRS spot height of $0.5$\,mm would result in all of the experimental desorption amounts falling within twice the uncertainty. We are confident that this error estimate also accounts for other small sources of uncertainty, e.g. the FEL-2 irradiation energy varying $\sim 5\%$ during irradiations and IR band strengths variations for pure CO vs. CO in contact with PHP, both of which are difficult to independently quantify.

When this uncertainty is taken into account, and after correcting with expected CO deposition for the layered ice, only excitation of the PHP mode leads to a decrease in $N_\mathrm{CO}$. 
Interestingly, for the layered ice, the desorbed column density corresponds to $86\%$ of the available CO, indicating that almost all the CO molecules within the irradiation spot desorbed during the irradiation. Conversely, for the mixed ice, only $24\%$ of the CO desorbed, with much of the remaining CO molecules likely trapped within the matrix of PHP molecules. In this case, the continuous slow desorption observed in the QMS signal might result from CO molecules slowly diffusing out from within the PHP film. 

The total amount of CO desorbed from the layered ice is three times larger than for the mixed ice. This is consistent with the integrated desorption signal for the QMS data. The QMS data also revealed that the desorption for the layered ice occurred over a significantly shorter time. The efficiency of the photodesorption process, taking the different photon fluences into account, can be described in terms of a yield, $Y_\mathrm{des,inc}$, in CO molecules per incident photon, determined according to
\begin{equation} \label{eq:Yinc}
    Y_\mathrm{des,inc} = \frac{\Delta N_\mathrm{CO}}{\phi\ t},
\end{equation}
where $\phi$ is the photon flux and $t$ is the effective irradiation time.
As can be seen from Fig. \ref{fig:QMSmain}, the desorption processes did not continue significantly for the full $1$\,min irradiation period. The time constants obtained from the bi-exponential fits in Fig. \ref{fig:QMSfit} indicate that the slow desorption process reaches $5\%$ of its initial amplitude after $13$\,s for the mixed ice and $3$\,s for the layered ice. We take these to be the effective irradiation times from which to derive the incident fluence. 
Alternatively, the desorption yield can be considered with respect to the absorbed photon fluence, $Y_\mathrm{des,abs}$, also referred to as the quantum yield, determined through
\begin{equation} \label{eq:Yabs}
    Y_\mathrm{des,abs} = \frac{Y_\mathrm{des,inc}}{\chi},
\end{equation}
where $\chi = 1-10^{-\mathrm{Abs}}$ is the fraction of the incident photons that are absorbed by the ice. This can be determined from the experimental absorbance, $\mathrm{Abs}$, at the relevant wavelength and spot on the sample using the Beer-Lambert law.
We note that this should be considered an estimate given the difference between using the unpolarized RAIRS beam with an incidence angle of $80\degree$ and the $p$-polarized light FEL beam with an incidence angle of $45\degree$. We do not take the spectral width of the FEL beam into account because the full FEL beam width is always contained within the extent of the PHP CH stretching band.
Table \ref{tab:results} summarises the resulting desorption amounts and yields for irradiation of the PHP band for both ices investigated, along with the effective irradiation time used to obtain the yields. The obtained yields indicate that it is an order of magnitude more efficient to desorb CO from the layered than the mixed ice through excitation of the PHP mode.

The quantum yields reported here can be compared to similar values derived from the desorption cross-sections in Table~\ref{tab:QMSfit}, although only for the slow process since reliable cross-sections could not be obtained for the fast process. These are determined using photon absorption cross-sections found from the experimental absorbance similar to $\chi$. The resulting values are $(0.3\pm0.1)\times 10^{-4}$ molecules per absorbed photon for the mixed ice and $(1.04\pm0.08)\times 10^{-4}$ molecules per absorbed photon for the layered ice. For the mixed ice, the correspondence with the quantum yield obtained from the RAIRS data confirms that, in this case, the slow process governs the majority of the desorption process. For the layered ice, the obtained quantum yield is around a third of the RAIRS derived value, reflecting the dominance of the fast process for this system.

The desorption yields listed in Table~\ref{tab:results} can be considered averages over the irradiation times. For the first-order desorption processes we observe in the studied ices, the desorption rate at any given time appears to depend on the amount of CO available on the surface. As CO desorbs, the desorption rate will therefore decrease. 
Further measurements are necessary to determine whether this also applies to other ice thicknesses.  
Under interstellar conditions, the timescale for the desorption process would be much longer, with continual replenishment of desorbed CO, maintaining a higher desorption rate, resulting in a higher overall yield.

\section{Discussion}

\subsection{Excitation of the CO mode}

Our results clearly indicate that IR excitation of the strong CO stretching mode at $4.67$\,\textmu m is not an efficient pathway by which CO can desorb from interstellar icy grains. No desorption at all was observed for the layered ice where CO was deposited as a thin film on top of PHP. This system can be considered analogous to pure CO on top of a hydrogenated carbonaceous grain. 
When the two species are mixed, a negligible amount of CO desorption is evident, indicating that the irradiation does excite the CO molecules. Similarly, we demonstrated in our previous work that excitation of CO for a pure CO ice and CO on top of ASW leads to no or inefficient desorption of CO \citep{Slumstrup2025} where we discussed how theoretical studies of CO support a low desorption efficiency \citep{Hemert2015,Ferrari2024}. 

The inefficient desorption observed following excitation of the CO stretching mode indicates that there must be some other non-desorptive energy dissipation pathway. Some relaxation could occur through coupling to the underlying substrate, although it is unclear whether this could be efficient for layers far away from the interface, especially considering that this would need to occur through the PHP layer. 
Alternatively, we suggest that the absorbed energy could be re-emitted through IR emission, as we also discussed for the CO and CO on ASW systems \citep{Slumstrup2025}. While this could not be measured at the LISA setup in its current configuration, \citet{Chen2019_Science} have demonstrated such a mechanism for a monolayer of CO adsorbed on a crystalline NaCl surface. They found that IR excitation of the CO led to vibrational energy pooling and subsequent IR emission from excited states.
It is unclear to what extent such a relaxation mechanism would also occur for the CO in our mixed and layered ices with PHP, although subsequent studies by \citet{DeVine2022} revealed that energy pooling and IR emission also occur for thick ($\sim 300$ layers) CO ices on NaCl. 
Re-emission may serve as a mechanism for dissipating energy from the ice systems following CO mode excitation. \citet{DeVine2022} showed that impurities in the CO ice could disrupt this process. In our experiments, PHP could act as such an impurity, opening up a pathway to CO desorption. 
We note that the small amount of CO desorption we observe for this ice could feasibly also arise from multiphoton effects. While this is unlikely under our experimental conditions, they cannot be completely ruled out, as discussed in Appendix \ref{App:PhotonDensity}.
Further investigations are necessary to verify the nature of the energy dissipation mechanism for vibrationally excited CO molecules in ices such as those studied in this work.

\subsection{Excitation of the PHP mode}

Compared to excitation of the CO stretching mode, our results show that vibrational excitation of PHP molecules deposited beneath or embedded within the CO ice leads to significant CO desorption at wavelengths where CO itself is transparent.
The desorption yields in Table \ref{tab:results} reveal that the resulting desorption is about an order of magnitude more efficient for the layered than for the mixed ice. For the layered ice, $86\%$ of the available CO molecules desorb, while for the mixed ice only $24\%$ desorb, with much of the remaining CO likely trapped within the PHP film. The gradual diffusion of these trapped CO molecules out from the PHP could contribute to the slower desorption process observed for the mixed ice.
Given the efficiency of the CO desorption for the layered ice with about 18 layers of CO, our results indicate that for a CO molecule to desorb, it does not need to be in direct contact with an excited PHP molecule. Rather, energy transfer must occur through the CO film to the vacuum interface, from where surface-bound CO molecules can desorb. It is noteworthy that this occurs despite the direct CO excitation not causing significant desorption. 
Collective effects due to PHP molecules being in contact with each other could also be responsible for the increased desorption efficiency in the case of the layered ice. As discussed in Appendix \ref{App:PhotonDensity}, we estimate that up to a third of the PHP molecules could be excited in each FEL micropulse, making it more likely for neighbouring PHP molecules to be excited simultaneously. Such effects would be much less likely in the mixed ice, where PHP molecules are dispersed.
It should be noted that the desorption yields reported in Table \ref{tab:results} are, in general, applicable only to the specific ice systems and thicknesses probed in these experiments. 
For example, a thicker PHP film would absorb more energy from the FEL, which could lead to a higher CO desorption rate, while a thicker CO film could either promote or hinder desorption, depending on the underlying desorption mechanism.
Further investigations are required to probe the dependence of the desorption efficiency on parameters such as layer thickness, concentration, and irradiation energies. We note that, under interstellar conditions, other processes such as UV excitation and the continuous replenishing of the ice from the gas phase will also occur simultaneously.
Nevertheless, we consider that our reported desorption yields provide a reasonable order of magnitude estimate for desorption from CO-PHP systems. 

The CO desorption efficiencies for the two CO-PHP systems studied can be compared to those we reported previously for the CO on ASW system \citep{Slumstrup2025}. For this layered ice, between $1.7\times 10^{-4}$ and $10.5\times 10^{-4}$ CO molecules desorbed per photon absorbed by the underlying ASW, with the yield increasing with photon energy. While the CO desorption efficiency for the mixed CO:PHP ice of $0.3\times 10^{-4}$ CO molecules per absorbed photon is an order of magnitude lower, the efficiency for the CO/PHP layered ice of $3.1\times 10^{-4}$ CO molecules per absorbed photon is within the range of that obtained with ASW.
The similarity in CO desorption efficiency for the two layered systems suggests that similar underlying mechanisms might contribute to the energy transfer to the CO ice, leading to CO desorption, despite the very different underlying molecules. Further investigations are required to fully determine the nature of these mechanisms.

For the CO on ASW system in \citet{Slumstrup2025}, a linear dependence between CO desorption and energy absorbed in the ASW indicates that the desorption is a single-photon process. 
As possible desorption mechanisms, we suggested excitation of the CO-\ce{H2O} intermolecular vibrations, local heating, and restructuring of the porous ASW, leading to a loss of strong CO binding sites. The porous ASW in the layered ice was observed to restructure similarly to pure porous ASW upon IR excitation, attributed to local reorganisation towards a more ordered structure \citep{Noble2020_a, Cuppen2022_ASW}.
For the CO-PHP systems in the present study, restructuring is less likely to cause significant loss of binding sites, as the IR spectrum in Fig. \ref{fig:IRdiff}(b) reveals the PHP layer to be non-porous. As discussed, there are also no significant spectral changes to the PHP bands indicating restructuring. 
During all irradiations, we observed a small temperature increase of ${\sim}\,0.1$\,K as measured by the Si-diode attached to the sample plate. We note that this does not directly reflect an actual temperature rise in the ice film under the laser spot but nevertheless indicates some heating of the ice. The temperature increase was approximately the same for all irradiation wavelengths, both on- and off-resonance, indicating that it likely represents energy input resulting from a small absorption by the Au substrate. This could not be responsible for the CO desorption in this work as it is seen only for on-resonance irradiations. For the CO-ASW system, temperature increases of up to $0.4$\,K were seen when exciting the strongest \ce{H2O} bands so some localised heating could be occurring. 
In the case of the CO-ASW system, there are overlapping vibrational modes, representing a possible pathway for energy transfer between the two species. It is clear from the spectrum in Fig. \ref{fig:IRSpectrum} that there is no such overlap for the CO-PHP systems in the mid-IR region. However, we speculate that phonon modes at longer wavelengths could provide a means for similar energy transfer. Phonon-mediated long-range energy transfer could explain how CO desorption occurs for the 18 layers of CO, upon excitation of the underlying PHP.
The QMS data suggest the presence of two desorption channels with different cross-sections, indicating that multiple mechanisms contribute to the observed CO desorption. Theoretical investigations of energy transfer in such systems will provide valuable insight into their underlying mechanisms. 

This work, along with our previous study of the CO on ASW system \citep{Slumstrup2025}, shows that efficient energy transfer and indirect CO desorption can take place for several ice systems.
IR-induced desorption has also been shown to occur for \ce{H2O} \citep{Noble2020_a, Krasnopoler1998,Focsa2003,Slumstrup2025}, for CO from CO-ASW mixed ices upon excitation of \ce{H2O} vibrational modes \citep{Ingman2023}, and for CO from CO-\ce{CH3OH} (methanol) mixed ices upon excitation of \ce{CH3OH} vibrational modes \citep{Santos2023}. These findings indicate that IR-induced non-thermal desorption processes could play a key role in determining the balance between the gas and condensed phases of the ISM.

\section{Astrophysical implications}

Our results confirm that direct vibrational excitation of the stretching mode of CO does not efficiently lead to CO desorption, indicating the presence of another energy dissipation pathway for excited CO molecules in ices. The possibility for dissipation through IR emission warrants further investigation in order to determine its impact on energy dissipation in vibrationally excited astrophysical ices. Understanding the effect of impurities in the ice on these processes is also important when considering astrophysically relevant ices, particularly when their presence might open up additional photodesorption pathways as indicated by this study. 

Our measurements indicate that IR photon absorption by PHP drives efficient CO desorption. Given the structural similarities between PHP and the broader class of interstellar carbonaceous materials, including PAHs, HPAHs, and hydrogenated carbonaceous grains, we suggest that they could lead to similar desorption processes. These materials are abundant in interstellar clouds and possess a range of strong IR bands. Recent observations of a range of cyano-PAHs in similar abundance in the dense cloud TMC-1 \citep{Wenzel2025b} suggest that PAHs and related species -- upon IR excitation and regardless of size or structure -- could contribute in a similar way to desorbing CO and potentially other small volatiles from icy grains. 
Carbonaceous species could, therefore, provide new energy input pathways to the ices where volatile species themselves are transparent or do not desorb due to non-desorptive relaxation processes, such as the IR emission observed in the case of CO. Such desorption processes provide a crucial link between condensed and gas-phase interstellar chemistry. 
In the mixed ice, PHP acts as an analogue of (H)PAHs embedded in CO ices with the relevance of such a system supported by the model for PAH freeze-out in dense clouds by \citet{Bouwman2011_II}.
In the layered ice, the PHP acts, instead, as an analogue of hydrogenated carbonaceous grains, thought to be abundant in the ISM \citep{Jones2016_III, Sandford1991, Pendleton1994, Chiar2000, Chiar2021}. This system should be considered as a general example of grain-induced desorption of volatile species. \ce{H2O} ice is expected to form first on the interstellar grains, but we note that the potential of bare grain surfaces being exposed to CO ice is supported by studies suggesting that \ce{H2O} could be mobile and cluster on the grain surfaces in dense clouds \citep{Rosu-Finsen2016} and by studies arguing that interstellar grains could be very porous and thus have low effective ice coverage \citep{Potapov2020}. 

It is necessary to consider the different photon fluxes present within interstellar clouds to relate our experimental results to astrophysical environments. The IR photon flux at low extinctions ($A_\mathrm{V}\, {\sim}\, 3$~mag), when CO has begun to form a nonpolar ice layer on the grains, has been estimated to be on the order of $10^{10}$\,photons\,cm$^{-2}$\,s$^{-1}$, across the $1$\,-\,$10$\,\textmu m range \citep{Boogert2015,Mathis1983}. At higher extinctions, deeper within the clouds ($A_\mathrm{V}\, {\sim}\, 50$~mag), where catastrophic freeze-out of CO has occurred and icy mantles are enriched with more complex molecules, the corresponding IR fluxes are on the order of $10^8$\,photons\,cm$^{-2}$\,s$^{-1}$.
Under our experimental conditions, typical photon fluences are on the order of $10^{22}$\,photons\,cm$^{-2}$, meaning that the derived photodesorption yields correspond to an interstellar desorption timescale of $10^4$\,-\,$10^6$\,years, depending on cloud density. This is within typical molecular cloud lifetimes, meaning that the experimental photon fluences are not unreasonable compared to those expected to impinge on an interstellar icy grain.
This indicates that the non-thermal desorption mechanisms discussed could play a significant role in gas-grain interactions.

Another relevant comparison is between the relative importance of IR and UV radiation in driving desorption. Several experimental studies have shown that UV radiation can also drive photodesorption of CO. For pure CO ice, UV and VUV photodesorption yields were found using broadband sources \citep{Oberg2009, Paardekooper2016} to be ${\sim}\,1\times10^{-3}$\,CO\,molecules/incident\,photon or using monochromatic VUV synchrotron radiation to be up to $5\times10^{-2}$\,CO\, molecules/incident\,photon for wavelengths corresponding to electronic transitions to the $\mathrm{A^1\Pi}$ state \citep{Fayolle2011}.
These are 3-4 orders of magnitudes higher than those obtained in the present work for IR excitations of the PHP mode. The flux of IR photons inside dense clouds is, however, much higher than for UV.
The integrated UV photon flux has been estimated to be on the order of $10^4$\,photons\,cm$^{-2}$\,s$^{-1}$ inside interstellar clouds \citep{Cecchi-Pestellini1992, Shen2004} for extinctions higher than $A_\mathrm{V}\,{\sim}\, 3$~mag, i.e. several orders of magnitude lower than the IR flux across the full wavelength range.
Hence, under dense cloud conditions, where IR photons dominate, such IR-induced processes could play a significantly larger role in photodesorption inside dense clouds.

We note that the quantitative extrapolations of our results to interstellar conditions are tentative, since we cannot fully rule out cumulative effects from excitations of neighbouring PHP molecules, an effect unlikely to occur with the significantly lower photon fluxes in the ISM.

\section{Conclusions}
This work demonstrates that (i) direct vibrational excitation of CO does not efficiently drive CO desorption, while (ii) vibrational excitation of C-H stretching modes in PHP leads to significant indirect photodesorption of CO.
We therefore suggest that exposure of carbonaceous materials containing C-H bonds in dense clouds to IR radiation could play a significant role in the desorption of volatile species from the grains. Further studies are required to understand the energy transfer and desorption mechanisms, as well as which ice constituents, thicknesses, and volatile species it occurs for.
With the abundance of (H)PAHs, hydrogenated grains, and IR irradiation in dense clouds and icy mantles, and the need to understand the existence of observed volatile species in the gas phase, we stress the importance of investigating and then including these desorption processes in astrochemical modelling.

\begin{acknowledgements}
This work has been supported by the Danish National Research Foundation through the Center of Excellence ``InterCat'' (Grant agreement no.: DNRF150). This project has received funding from the European Union’s Horizon 2020 research and innovation programme under grant agreement no. 654148 - LASERLAB-EUROPE. 

The main components of the experimental apparatus LISA used in this work were purchased using funding obtained from the Royal Society through grants UF130409, RGF/EA/180306, and URF/R/191018. The authors thank the FELIX Laboratory team for their experimental assistance and support.

\end{acknowledgements}

\bibliographystyle{aa} 
\bibliography{bibliography}

\begin{thebibliography}{81}
\expandafter\ifx\csname natexlab\endcsname\relax\def\natexlab#1{#1}\fi

\bibitem[{Basalg{\`{e}}te {et~al.}(2021)Basalg{\`{e}}te, Oca{\~{n}}a, F{\'{e}}raud, Romanzin, Philippe, Michaut, Fillion, \& Bertin}]{Basalgete2021}
Basalg{\`{e}}te, R., Oca{\~{n}}a, A.~J., F{\'{e}}raud, G., {et~al.} 2021, Astrophys. J., 922, 213

\bibitem[{Bernstein {et~al.}(2003)Bernstein, Moore, Elsila, Sandford, Allamandola, \& Zare}]{Bernstein2003}
Bernstein, M.~P., Moore, M.~H., Elsila, J.~E., {et~al.} 2003, Astrophys. J., 582, L25

\bibitem[{Bernstein {et~al.}(1999)Bernstein, Sandford, Allamandola, Gillette, Clemett, \& Zare}]{Bernstein1999}
Bernstein, M.~P., Sandford, S.~A., Allamandola, L.~J., {et~al.} 1999, Science, 283, 1135

\bibitem[{Bertin {et~al.}(2012)Bertin, Fayolle, Romanzin, {\"{O}}berg, Michaut, Moudens, Philippe, Jeseck, Linnartz, \& Fillion}]{Bertin2012}
Bertin, M., Fayolle, E.~C., Romanzin, C., {et~al.} 2012, Phys. Chem. Chem. Phys., 14, 9929

\bibitem[{Bertin {et~al.}(2016)Bertin, Romanzin, Doronin, Philippe, Jeseck, Ligterink, Linnartz, Michaut, \& Fillion}]{Bertin2016}
Bertin, M., Romanzin, C., Doronin, M., {et~al.} 2016, Astrophys. J. Lett., 817, L12

\bibitem[{Boogert {et~al.}(2015)Boogert, Gerakines, \& Whittet}]{Boogert2015}
Boogert, A. C.~A., Gerakines, P.~A., \& Whittet, D. C.~B. 2015, Annu. Rev. Astron. Astrophys., 53, 541

\bibitem[{Bouilloud {et~al.}(2015)Bouilloud, Fray, Cottin, Gazeau, \& Jolly}]{Bouilloud2015}
Bouilloud, M., Fray, N., Cottin, H., Gazeau, M., \& Jolly, A. 2015, Mon. Not. R. Astron. Soc., 451, 2145

\bibitem[{Bouwman {et~al.}(2011{\natexlab{a}})Bouwman, Cuppen, Steglich, Allamandola, \& Linnartz}]{Bouwman2011_II}
Bouwman, J., Cuppen, H.~M., Steglich, M., Allamandola, L.~J., \& Linnartz, H. 2011{\natexlab{a}}, Astron. Astrophys., 529, A46

\bibitem[{Bouwman {et~al.}(2011{\natexlab{b}})Bouwman, Mattioda, Linnartz, \& Allamandola}]{Bouwman2011_I}
Bouwman, J., Mattioda, A.~L., Linnartz, H., \& Allamandola, L.~J. 2011{\natexlab{b}}, Astron. Astrophys., 525, A93

\bibitem[{Cazaux {et~al.}(2016)Cazaux, Boschman, Rougeau, Reitsma, Hoekstra, Teillet-Billy, Morisset, Spaans, \& Schlath{\"{o}}lter}]{Cazaux2016}
Cazaux, S., Boschman, L., Rougeau, N., {et~al.} 2016, Sci. Rep., 6, 19835

\bibitem[{Cecchi-Pestellini \& Aiello(1992)}]{Cecchi-Pestellini1992}
Cecchi-Pestellini, C. \& Aiello, S. 1992, Mon. Not. R. Astron. Soc., 258, 125

\bibitem[{Cernicharo {et~al.}(2012)Cernicharo, Marcelino, Roueff, Gerin, Jim{\'{e}}nez-Escobar, \& {Mu{\~{n}}oz Caro}}]{Cernicharo2012}
Cernicharo, J., Marcelino, N., Roueff, E., {et~al.} 2012, Astrophys. J. Lett., 759, 2010

\bibitem[{Chen {et~al.}(2019)Chen, Lau, Schwarzer, Verma, \& Wodtke}]{Chen2019_Science}
Chen, L., Lau, J.~A., Schwarzer, D., Verma, V.~B., \& Wodtke, A.~M. 2019, Science, 363, 158

\bibitem[{Chiar {et~al.}(2021)Chiar, de~Barros, Mattioda, \& Ricca}]{Chiar2021}
Chiar, J.~E., de~Barros, A. L.~F., Mattioda, A.~L., \& Ricca, A. 2021, Astrophys. J., 908, 239

\bibitem[{Chiar {et~al.}(2000)Chiar, Tielens, Whittet, Schutte, Boogert, Lutz, van Dishoeck, \& Bernstein}]{Chiar2000}
Chiar, J.~E., Tielens, A. G. G.~M., Whittet, D. C.~B., {et~al.} 2000, Astrophys. J., 537, 749

\bibitem[{Cook {et~al.}(2015)Cook, Ricca, Mattioda, Bouwman, Roser, Linnartz, Bregman, \& Allamandola}]{Cook2015}
Cook, A.~M., Ricca, A., Mattioda, A.~L., {et~al.} 2015, Astrophys. J., 799, 14

\bibitem[{Coussan {et~al.}(2022)Coussan, Noble, Cuppen, Redlich, \& Ioppolo}]{Coussan2022}
Coussan, S., Noble, J.~A., Cuppen, H.~M., Redlich, B., \& Ioppolo, S. 2022, J. Phys. Chem. A, 126, 2262

\bibitem[{Coussan {et~al.}(2015)Coussan, Roubin, \& Noble}]{Coussan2015}
Coussan, S., Roubin, P., \& Noble, J.~A. 2015, Phys. Chem. Chem. Phys., 17, 9429

\bibitem[{Cuppen {et~al.}(2022)Cuppen, Noble, Coussan, Redlich, \& Ioppolo}]{Cuppen2022_ASW}
Cuppen, H.~M., Noble, J.~A., Coussan, S., Redlich, B., \& Ioppolo, S. 2022, J. Phys. Chem. A, 126, 8859

\bibitem[{Cuylle {et~al.}(2014)Cuylle, Allamandola, \& Linnartz}]{Cuylle2014}
Cuylle, S.~H., Allamandola, L.~J., \& Linnartz, H. 2014, Astron. Astrophys., 562

\bibitem[{DeVine {et~al.}(2022)DeVine, Choudhury, Lau, Schwarzer, \& Wodtke}]{DeVine2022}
DeVine, J.~A., Choudhury, A., Lau, J.~A., Schwarzer, D., \& Wodtke, A.~M. 2022, J. Phys. Chem. A, 126, 2270

\bibitem[{Dupuy {et~al.}(2017)Dupuy, Bertin, F{\'{e}}raud, Michaut, Jeseck, Doronin, Philippe, Romanzin, \& Fillion}]{Dupuy2017}
Dupuy, R., Bertin, M., F{\'{e}}raud, G., {et~al.} 2017, Astron. Astrophys., 603, A61

\bibitem[{Evans {et~al.}(2025)Evans, Booth, Walsh, Ilee, Keyte, Law, Leemker, Notsu, {\"{O}}berg, Temmink, \& van~der Marel}]{Evans2025}
Evans, L., Booth, A.~S., Walsh, C., {et~al.} 2025, Astrophys. J., 982, 62

\bibitem[{Fayolle {et~al.}(2011)Fayolle, Bertin, Romanzin, Michaut, \& Oberg}]{Fayolle2011}
Fayolle, E.~C., Bertin, M., Romanzin, C., Michaut, X., \& Oberg, K.~I. 2011, Astrophys. J. Lett., 739, 1

\bibitem[{Fedoseev {et~al.}(2017)Fedoseev, Chuang, Ioppolo, Qasim, van Dishoeck, \& Linnartz}]{Fedoseev2017}
Fedoseev, G., Chuang, K.~J., Ioppolo, S., {et~al.} 2017, Astrophys. J., 842, 52

\bibitem[{Ferrari {et~al.}(2024)Ferrari, Hemert, Meyer, \& Lamberts}]{Ferrari2024}
Ferrari, B.~C., Hemert, M.~V., Meyer, J., \& Lamberts, T. 2024, J. Phys. Chem. C, 128, 21060

\bibitem[{Focsa {et~al.}(2003)Focsa, Chazallon, \& Destombes}]{Focsa2003}
Focsa, C., Chazallon, B., \& Destombes, J.~L. 2003, Surf. Sci., 528, 189

\bibitem[{Fuchs {et~al.}(2009)Fuchs, Cuppen, Ioppolo, Romanzin, Bisschop, Andersson, van Dishoeck, \& Linnartz}]{Fuchs2009}
Fuchs, G.~W., Cuppen, H.~M., Ioppolo, S., {et~al.} 2009, Astron. Astrophys., 505, 629

\bibitem[{Gudipati \& Yang(2012)}]{Gudipati2012}
Gudipati, M.~S. \& Yang, R. 2012, Astrophys. J. Lett., 756, L24

\bibitem[{Guzm{\'{a}}n {et~al.}(2013)Guzm{\'{a}}n, Goicoechea, Pety, Gratier, Gerin, Roueff, {Le Petit}, {Le Bourlot}, \& Faure}]{Guzman2013}
Guzm{\'{a}}n, V.~V., Goicoechea, J.~R., Pety, J., {et~al.} 2013, Astron. Astrophys., 560, A73

\bibitem[{Habart {et~al.}(2003)Habart, Boulanger, Verstraete, {Pineau des For{\^{e}}ts}, Falgarone, \& Abergel}]{Habart2003}
Habart, E., Boulanger, F., Verstraete, L., {et~al.} 2003, Astron. Astrophys., 397, 623

\bibitem[{Hardegree-Ullman {et~al.}(2014)Hardegree-Ullman, Gudipati, Boogert, Lignell, Allamandola, Stapelfeldt, \& Werner}]{Hardegree-Ullman2014}
Hardegree-Ullman, E.~E., Gudipati, M.~S., Boogert, A.~C., {et~al.} 2014, Astrophys. J., 784, 172

\bibitem[{Hemert {et~al.}(2015)Hemert, Takahashi, \& Dishoeck}]{Hemert2015}
Hemert, M. C.~V., Takahashi, J., \& Dishoeck, E. F.~V. 2015, J. Phys. Chem. A, 119, 6354

\bibitem[{Ingman {et~al.}(2023)Ingman, Laurinavicius, Zhang, Schrauwen, Redlich, Noble, Ioppolo, McCoustra, \& Brown}]{Ingman2023}
Ingman, E.~R., Laurinavicius, D., Zhang, J., {et~al.} 2023, Faraday Discuss., 245, 446

\bibitem[{Ioppolo {et~al.}(2021)Ioppolo, Fedoseev, Chuang, Cuppen, Clements, Jin, Garrod, Qasim, Kofman, van Dishoeck, \& Linnartz}]{Ioppolo2021}
Ioppolo, S., Fedoseev, G., Chuang, K.~J., {et~al.} 2021, Nat. Astron., 5, 197

\bibitem[{Ioppolo {et~al.}(2022)Ioppolo, Noble, {Traspas Mui{\~{n}}a}, Cuppen, Coussan, Redlich, Traspas, Cuppen, Coussan, \& Redlich}]{Ioppolo2022}
Ioppolo, S., Noble, J.~A., {Traspas Mui{\~{n}}a}, A., {et~al.} 2022, J. Mol. Spectrosc., 385, 111601

\bibitem[{Jensen {et~al.}(2019)Jensen, Leccese, Simonsen, Skov, Bonfanti, Thrower, Martinazzo, \& Hornek{\ae}r}]{Jensen2019}
Jensen, P.~A., Leccese, M., Simonsen, F. D.~S., {et~al.} 2019, Mon. Not. R. Astron. Soc., 486, 5492

\bibitem[{Joblin \& Tielens(2011)}]{JoblinTielens2011}
Joblin, C. \& Tielens, A. 2011, in EAS Publ. Ser., ed. C.~Joblin \& A.~G. G.~M. Tielens, Vol.~46

\bibitem[{Jones(2016)}]{Jones2016_III}
Jones, A.~P. 2016, R. Soc. Open Sci., 3, 160224

\bibitem[{Kouchi {et~al.}(2021)Kouchi, Tsuge, Hama, Niinomi, Nakatani, Shimonishi, Oba, Kimura, Sirono, Okuzumi, Momose, Furuya, \& Watanabe}]{Kouchi2021}
Kouchi, A., Tsuge, M., Hama, T., {et~al.} 2021, Mon. Not. R. Astron. Soc., 505, 1530

\bibitem[{Krasnopoler \& George(1998)}]{Krasnopoler1998}
Krasnopoler, A. \& George, S.~M. 1998, J. Phys. Chem. B, 102, 788

\bibitem[{Mathis {et~al.}(1983)Mathis, Mezger, \& Panagia}]{Mathis1983}
Mathis, J.~S., Mezger, P.~G., \& Panagia, N. 1983, Astron. Astrophys., 128, 212

\bibitem[{Meinert {et~al.}(2016)Meinert, Myrgorodska, de~Marcellus, Buhse, Nahon, Hoffmann, D'Hendecourt, \& Meierhenrich}]{Meinert2016}
Meinert, C., Myrgorodska, I., de~Marcellus, P., {et~al.} 2016, Science, 352, 208

\bibitem[{Mennella {et~al.}(2012)Mennella, Hornek{\ae}r, Thrower, \& Accolla}]{Mennella2012}
Mennella, V., Hornek{\ae}r, L., Thrower, J., \& Accolla, M. 2012, Astrophys. J. Lett., 745, L2

\bibitem[{{Mu{\~{n}}oz Caro} {et~al.}(2002){Mu{\~{n}}oz Caro}, Meierhenrich, Schutte, Barbier, Segovia, Rosenbauer, Thiemann, Brack, \& Greenberg}]{MunozCaro2002}
{Mu{\~{n}}oz Caro}, G., Meierhenrich, U., Schutte, W., {et~al.} 2002, Nature, 416, 403

\bibitem[{Naraoka {et~al.}(2023)Naraoka, Takano, Dworkin, Oba, Hamase, Furusho, Ogawa, Hashiguchi, Fukushima, Aoki, Schmitt-Kopplin, Aponte, Parker, Glavin, McLain, Elsila, Graham, Eiler, Orthous-Daunay, Wolters, Isa, Vuitton, Thissen, Sakai, Yoshimura, Koga, Ohkouchi, Chikaraishi, Sugahara, Mita, Furukawa, Hertkorn, Ruf, Yurimoto, Nakamura, Noguchi, Okazaki, Yabuta, Sakamoto, Tachibana, Connolly, Lauretta, Abe, Yada, Nishimura, Yogata, Nakato, Yoshitake, Suzuki, Miyazaki, Furuya, Hatakeda, Soejima, Hitomi, Kumagai, Usui, Hayashi, Yamamoto, Fukai, Kitazato, Sugita, Namiki, Arakawa, Ikeda, Ishiguro, Hirata, Wada, Ishihara, Noguchi, Morota, Sakatani, Matsumoto, Senshu, Honda, Tatsumi, Yokota, Honda, Michikami, Matsuoka, Miura, Noda, Yamada, Yoshihara, Kawahara, Ozaki, Iijima, Yano, Hayakawa, Iwata, Tsukizaki, Sawada, Hosoda, Ogawa, Okamoto, Hirata, Shirai, Shimaki, Yamada, Okada, Yamamoto, Takeuchi, Fujii, Takei, Yoshikawa, Mimasu, Ono, Ogawa, Kikuchi, Nakazawa, Terui, Tanaka, Saiki, Yoshikawa, Watanabe, \&
  Tsuda}]{Naraoka2023}
Naraoka, H., Takano, Y., Dworkin, J.~P., {et~al.} 2023, Science, 379, eabn9033

\bibitem[{Noble {et~al.}(2020{\natexlab{a}})Noble, Cuppen, Coussan, Redlich, \& Ioppolo}]{Noble2020_a}
Noble, J.~A., Cuppen, H.~M., Coussan, S., Redlich, B., \& Ioppolo, S. 2020{\natexlab{a}}, J. Phys. Chem. C, 124, 20864

\bibitem[{Noble {et~al.}(2014{\natexlab{a}})Noble, Martin, Fraser, Roubin, \& Coussan}]{Noble2014a}
Noble, J.~A., Martin, C., Fraser, H.~J., Roubin, P., \& Coussan, S. 2014{\natexlab{a}}, J. Phys. Chem. C, 118, 20488

\bibitem[{Noble {et~al.}(2014{\natexlab{b}})Noble, Martin, Fraser, Roubin, \& Coussan}]{Noble2014b}
Noble, J.~A., Martin, C., Fraser, H.~J., Roubin, P., \& Coussan, S. 2014{\natexlab{b}}, J. Phys. Chem. Lett., 5, 826

\bibitem[{Noble {et~al.}(2020{\natexlab{b}})Noble, Michoulier, Aupetit, \& Mascetti}]{Noble2020_b}
Noble, J.~A., Michoulier, E., Aupetit, C., \& Mascetti, J. 2020{\natexlab{b}}, Astron. Astrophys., 644, A22

\bibitem[{Nuevo {et~al.}(2011)Nuevo, Milam, Sandford, {De Gregorio}, Cody, \& Kilcoyne}]{Nuevo2011}
Nuevo, M., Milam, S.~N., Sandford, S.~A., {et~al.} 2011, Adv. Sp. Res., 48, 1126

\bibitem[{{\"{O}}berg(2016)}]{Oberg2016}
{\"{O}}berg, K.~I. 2016, Chem. Rev., 116, 9631

\bibitem[{{\"{O}}berg {et~al.}(2015){\"{O}}berg, Furuya, Loomis, Aikawa, Andrews, Qi, Dishoeck, \& Wilner}]{Oberg2015}
{\"{O}}berg, K.~I., Furuya, K., Loomis, R., {et~al.} 2015, Astrophys. J., 810, 112

\bibitem[{{\"{O}}berg {et~al.}(2009){\"{O}}berg, van Dishoeck, \& Linnartz}]{Oberg2009}
{\"{O}}berg, K.~I., van Dishoeck, E.~F., \& Linnartz, H. 2009, Astron. Astrophys., 496, 281

\bibitem[{Paardekooper {et~al.}(2016)Paardekooper, Fedoseev, Riedo, \& Linnartz}]{Paardekooper2016}
Paardekooper, D.~M., Fedoseev, G., Riedo, A., \& Linnartz, H. 2016, Astron. Astrophys., 596, A72

\bibitem[{Pedretti {et~al.}(2021)Pedretti, Mazzolari, Gervasoni, Fumagalli, \& Vistoli}]{Pedretti2021_VEGAzz}
Pedretti, A., Mazzolari, A., Gervasoni, S., Fumagalli, L., \& Vistoli, G. 2021, Bioinformatics, 37, 1174

\bibitem[{Pendleton {et~al.}(1994)Pendleton, Sandford, Allamandola, Tielens, \& Sellgren}]{Pendleton1994}
Pendleton, Y.~J., Sandford, S.~A., Allamandola, L.~J., Tielens, A. G. G.~M., \& Sellgren, K. 1994, Astrophys. J., 437, 683

\bibitem[{Perotti {et~al.}(2020)Perotti, Rocha, J{\o}rgensen, Kristensen, Fraser, \& Pontoppidan}]{Perotti2020}
Perotti, G., Rocha, W.~R., J{\o}rgensen, J.~K., {et~al.} 2020, Astron. Astrophys., 643

\bibitem[{Pi{\'{e}}tu {et~al.}(2007)Pi{\'{e}}tu, Dutrey, \& Guilloteau}]{Pietu2007}
Pi{\'{e}}tu, V., Dutrey, A., \& Guilloteau, S. 2007, Astron. Astrophys., 467, 163

\bibitem[{Porter \& Strong(2005)}]{Porter2005}
Porter, T.~A. \& Strong, A.~W. 2005, in 29th Int. Cosm. Ray Conf.

\bibitem[{Potapov {et~al.}(2020)Potapov, J{\"{a}}ger, \& Henning}]{Potapov2020}
Potapov, A., J{\"{a}}ger, C., \& Henning, T. 2020, Phys. Rev. Lett., 124, 221103

\bibitem[{Rauls \& Hornek{\ae}r(2008)}]{RaulsHornekaer2008}
Rauls, E. \& Hornek{\ae}r, L. 2008, Astrophys. J., 679, 531

\bibitem[{Rosu-finsen {et~al.}(2016)Rosu-finsen, Marchione, Salter, Stubbing, Brown, \& Mccoustra}]{Rosu-Finsen2016}
Rosu-finsen, A., Marchione, D., Salter, T.~L., {et~al.} 2016, Phys.Chem.Chem.Phys., 18, 31930

\bibitem[{Roueff {et~al.}(2013)Roueff, Ruaud, {Le Petit}, Godard, \& {Le Bourlot}}]{Roueff2013}
Roueff, E., Ruaud, M., {Le Petit}, F., Godard, B., \& {Le Bourlot}, J. 2013, Proc. Int. Astron. Union, 9, 311

\bibitem[{Sandford {et~al.}(1991)Sandford, Allamandola, Tielens, Sellgren, Tapia, \& Pendleton}]{Sandford1991}
Sandford, S.~A., Allamandola, L.~J., Tielens, A. G. G.~M., {et~al.} 1991, Astrophys. J., 371, 607

\bibitem[{Sandford {et~al.}(2013)Sandford, Bernstein, \& Materese}]{Sandford2013}
Sandford, S.~A., Bernstein, M.~P., \& Materese, C.~K. 2013, Astrophys. Journal, Suppl. Ser., 205, 8

\bibitem[{Santos {et~al.}(2023)Santos, Chuang, Schrauwen, {Traspas Mui{\~{n}}a}, Zhang, Cuppen, Redlich, Linnartz, \& Ioppolo}]{Santos2023}
Santos, J.~C., Chuang, K.-J., Schrauwen, J. G.~M., {et~al.} 2023, Astron. Astrophys., 112, 1

\bibitem[{Schrauwen {et~al.}(2024)Schrauwen, Cuppen, Ioppolo, \& Redlich}]{Schrauwen2024}
Schrauwen, J. G.~M., Cuppen, H.~M., Ioppolo, S., \& Redlich, B. 2024, Astron. Astrophys., 691

\bibitem[{Sephton(2002)}]{Sephton2002}
Sephton, M.~A. 2002, Nat. Prod. Rep., 19, 292

\bibitem[{Shen {et~al.}(2004)Shen, Greenberg, Schutte, \& van Dishoeck}]{Shen2004}
Shen, C.~J., Greenberg, J.~M., Schutte, W.~A., \& van Dishoeck, E.~F. 2004, Astron. Astrophys., 415, 203

\bibitem[{Sloan {et~al.}(1997)Sloan, Bregman, Geballe, Allamandola, Woodward, \& Woodward1}]{Sloan1997}
Sloan, G.~C., Bregman, J.~D., Geballe, T.~R., {et~al.} 1997, Astrophys. J., 474, 735

\bibitem[{Slumstrup {et~al.}(2025)Slumstrup, Thrower, Schrauwen, Lamberts, Ingman, Laurinavicius, Devine, Scheltinga, Santos, Noble, Wenzel, Mccoustra, Brown, Linnartz, Hornek{\ae}r, Cuppen, Redlich, \& Ioppolo}]{Slumstrup2025}
Slumstrup, L., Thrower, J.~D., Schrauwen, J. G.~M., {et~al.} 2025, ACS Earth Sp. Chem., 9, 1607

\bibitem[{Steglich {et~al.}(2013)Steglich, Huisken, Friedrich, \& Plass}]{Steglich2013}
Steglich, M., Huisken, F., Friedrich, M., \& Plass, W. 2013, Astrophys. J. Suppl. Ser., 208, 26

\bibitem[{Thrower {et~al.}(2012)Thrower, J{\o}rgensen, Friis, Baouche, Mennella, Luntz, Andersen, Hammer, \& Hornek{\ae}r}]{Thrower2012}
Thrower, J.~D., J{\o}rgensen, B., Friis, E.~E., {et~al.} 2012, Astrophys. J., 752, 3

\bibitem[{Tielens(2008)}]{Tielens2008}
Tielens, A. 2008, Annu. Rev. Astron. Astrophys., 46, 289

\bibitem[{Vastel {et~al.}(2014)Vastel, Ceccarelli, Lefloch, \& Bachiller}]{Vastel2014}
Vastel, C., Ceccarelli, C., Lefloch, B., \& Bachiller, R. 2014, Astrophys. J. Lett., 795, 6

\bibitem[{Wenzel {et~al.}(2024)Wenzel, Cooke, Changala, Bergin, Zhang, Burkhardt, Byrne, Charnley, Cordiner, Duffy, Fried, Gupta, Holdren, Lipnicky, Loomis, Shay, Shingledecker, Siebert, Stewart, Willis, Xue, Remijan, Wendlandt, McCarthy, \& McGuire}]{Wenzel2024}
Wenzel, G., Cooke, I.~R., Changala, P.~B., {et~al.} 2024, Science, 386, 810

\bibitem[{Wenzel {et~al.}(2025{\natexlab{a}})Wenzel, Gong, Xue, Changala, Holdren, Speak, Stewart, Fried, Willis, Bergin, Burkhardt, Byrne, Charnley, Lipnicky, Loomis, Shingledecker, Cooke, McCarthy, Remijan, Wendlandt, \& McGuire}]{Wenzel2025b}
Wenzel, G., Gong, S., Xue, C., {et~al.} 2025{\natexlab{a}}, Astrophys. J. Lett., 984, L36

\bibitem[{Wenzel {et~al.}(2025{\natexlab{b}})Wenzel, Speak, Changala, Willis, Burkhardt, Zhang, Bergin, Byrne, Charnley, Fried, Gupta, Herbst, Holdren, Lipnicky, Loomis, Shingledecker, Xue, Remijan, Wendlandt, Mccarthy, Cooke, \& McGuire}]{Wenzel2025a}
Wenzel, G., Speak, T.~H., Changala, P.~B., {et~al.} 2025{\natexlab{b}}, Nat. Astron., 9, 262

\bibitem[{Willacy \& Langer(2000)}]{Willacy2000}
Willacy, K. \& Langer, W.~D. 2000, Astrophys. J., 544, 903

\bibitem[{Zeichner {et~al.}(2023)Zeichner, Aponte, Bhattacharjee, Dong, Hofmann, Dworkin, Glavin, Elsila, Graham, Naraoka, Takano, Tachibana, Karp, Grice, Holman, Freeman, Yurimoto, Nakamura, Noguchi, Okazaki, Yabuta, Sakamoto, Yada, Nishimura, Nakato, Miyazaki, Yogata, Abe, Okada, Usui, Yoshikawa, Saiki, Tanaka, Terui, Nakazawa, Watanabe, Tsuda, Hamase, Fukushima, Aoki, Hashiguchi, Mita, Chikaraishi, Ohkouchi, Ogawa, Sakai, Parker, McLain, Orthous-Daunay, Vuitton, Wolters, Schmitt-Kopplin, Hertkorn, Thissen, Ruf, Isa, Oba, Koga, Yoshimura, Araoka, Sugahara, Furusho, Furukawa, Aoki, Kano, Nomura, Sasaki, Sato, Yoshikawa, Tanaka, Morita, Onose, Kabashima, Fujishima, Yamazaki, Kimura, \& Eiler}]{Zeichner2023}
Zeichner, S.~S., Aponte, J.~C., Bhattacharjee, S., {et~al.} 2023, Science, 382, 1411

\end{thebibliography}

\begin{appendix}
\section{Spot size and overlap of RAIRS and FEL beams} \label{App:SpotAreas}
The spot sizes of the RAIRS and FEL beams incident on the substrate are both important factors for the analysis in this work. The FEL spot size, $S_\mathrm{FEL}$, determines the incident photon flux and also enters into the calculation of the desorbed CO column density, $\Delta N_\mathrm{CO}$. This calculation also requires the RAIRS spot size, $S_\mathrm{RAIRS}$, through the ratio $S_\mathrm{RAIRS}/S_\mathrm{FEL}$, which is referred to as the dilution factor. The details of the calculated spot sizes can be found in \citet{Slumstrup2025}.
For the RAIRS aperture of $1$\,mm used in this work, a spot height of $d_\mathrm{IR}=0.45$ cm was determined, resulting in $S_\mathrm{RAIRS}= 0.92$\,cm$^2$.
The FEL spot size is wavelength-dependent. 
For the irradiations in this work, FEL spot heights, areas, and dilution factors are reported in Table \ref{AppTab:spotsizesFEL}.

\begin{table}[h]
    \caption{FEL spot heights, areas, and dilution factors for the irradiations in this work.}
    \label{AppTab:spotsizesFEL}
    \centering
    \begin{tabular}{c c c c}
    \hline \hline
    $\lambda_\mathrm{FEL}$ ({\textmu}m) & $d_\mathrm{FEL}$ (mm) & $S_\mathrm{FEL}$ (cm$^2$) & $S_\mathrm{RAIRS}/S_\mathrm{FEL}$ \\
    \hline
    3.42 & 0.83 & 0.0077 & 120 \\
    4.02 & 0.97 & 0.0105 & 87 \\
    4.67 & 1.13 & 0.0143 & 64 \\
    \hline
    \end{tabular}
\end{table}

\section{PHP column and layer density} \label{App:PHPamounts}

To the best of our knowledge, there are no published IR band strengths for PHP. We therefore make an estimate by comparing IR and QMS measurements between PHP and pyrene for a series of temperature-programmed desorption (TPD) measurements at the LISA setup, as band strengths for pyrene exist in the literature. 
We assume the QMS to have similar sensitivity to pyrene ($m/z=202$) and PHP ($m/z=218$), so to a first approximation, equivalent pyrene and PHP signals in the QMS correspond to the same number of molecules. However, as we are only measuring the parent mass in both cases, we need to take the different fragmentation patterns into account. 
We therefore measured the fragmentation pattern of both molecules in the LISA QMS by dosing each molecule directly into the QMS and measuring all masses in the range $m/z=100-220$. We determined that for pyrene, $40.2\%$ is detected as the parent $m/z=202$, while for PHP, $17.5\%$ is detected as the parent $m/z=218$. By dividing the detected QMS signals for a parent mass by this detection efficiency, the QMS detection is corrected for the difference in detection for pyrene and PHP.
For three TPDs of PHP-containing ices, the integrated QMS desorption signal for $m/z=218$ was related to the signal for $m/z=202$ in an equivalent TPD of a pyrene-containing ice, after correcting for the QMS detection difference for each. This factor then represents the difference in pyrene/PHP amounts between measurements. 
The absolute pyrene amounts were estimated with eq. \eqref{eq:ColDens}, as an average of the $714$\,cm$^{-1}$ and $851$\,cm$^{-1}$ pyrene bands. We used band strengths for pyrene in \ce{H2O} \citep{Hardegree-Ullman2014} with $A_{714}=14.073\times 10^{-18}$ cm/molecule and $A_{851}=19.125\times 10^{-18}$ cm/molecule, respectively. 
Using the factor from the QMS comparison, the absolute PHP amount for each measurement was then estimated. This absolute amount was related to the integrated absorbance of the combined aliphatic CH stretching modes for PHP at $3.42$\,{\textmu}m to finally obtain an estimated band strength, through eq. \eqref{eq:ColDens}, of $A_\mathrm{CH}=1.53\times 10^{-16}$ cm/molecule.

The PHP molecule has a cross-sectional area of $3.79\times10^{-15}$\,cm$^2$, based on an estimate made in the molecular modelling suite VEGA ZZ \citep{Pedretti2021_VEGAzz} using the published molecular geometry for PHP\footnote{National Center for Biotechnology Information. PubChem Compound Summary for CID 75524, Perhydropyrene. Retrieved May 1, 2025 from \href{https://pubchem.ncbi.nlm.nih.gov/compound/Perhydropyrene}{\url{https://pubchem.ncbi.nlm.nih.gov/compound/Perhydropyrene}}}, resulting in a layer density for PHP molecules of $0.26\times10^{15}$\,cm$^{-2}$. Dividing the column density of PHP by this layer density yields the approximate number of PHP layers in the ice.

\section{Details of individual CO desorption peaks} \label{App:QMSpeaks}

\begin{figure}[ht]
    \centering
    \includegraphics{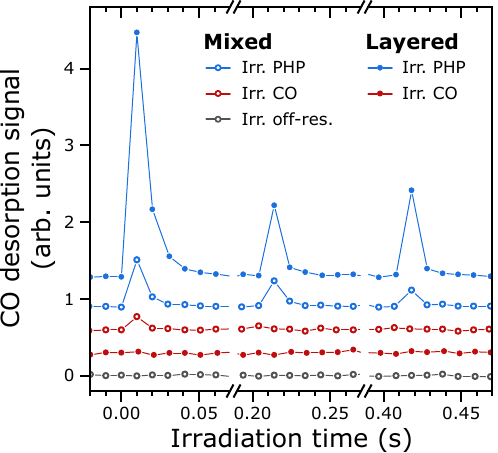}
    \caption{The CO ($m/z=28$) desorption detected in the QMS for the first few FEL macropulses during all irradiations in this work. The connecting lines are shown as a guide to the eye. All traces have been baselined to the average CO signal before irradiation starts and offset for clarity. The zero-time is defined as the onset of the first peak for the irradiations showing desorption signal.
    }
    \label{AppFig:QMSzoom}
\end{figure}

Figure \ref{AppFig:QMSzoom} shows the CO desorption signals in the QMS for all irradiations considered, focusing only on the desorption peaks corresponding to the first three FEL macropulses. Along with the CO desorption signals shown for PHP and CO irradiations in Fig. \ref{fig:QMSmain}, the off-resonance irradiation of the mixed ice is also included, confirming that no CO desorption occurs. 
In this figure, the data for the PHP irradiation of the layered ice has not been scaled as it was in Fig. \ref{fig:QMSmain}.

The temporal spacing of the desorption peaks reflects the $5$\,Hz macropulse repetition rate of the FEL while the acquisition bin size corresponds to the $10$\,ms QMS acquisition time.
In addition, there is a short processing period on the order of a few 10ths of a ms associated with each measurement point during which the QMS is not acquiring, and hence we do not expect to detect the true shape and height of each peak.
For the analysis and fits to the QMS signals in Fig. \ref{fig:QMSfit}, each macropulse peak has been integrated to account for as much of the desorption as possible, in a range of $100$\,ms around the peak.

We note that the broadening of the desorption transient associated with each macropulse does not arise from the much shorter $6$\,\textmu s width of the macropulse. Even a low kinetic temperature of $T_{k,\mathrm{CO}}=25$ K would result in a CO arrival time distribution at the QMS that would be contained within the $10$\,ms acquisition time. We attribute the broadening to the pump-down of CO in the UHV chamber following each desorption event.

\section{Estimate of absorbed photon density} \label{App:PhotonDensity}

The experimental photon fluxes in this work are orders of magnitudes higher than in the ISM. Therefore, before comparing the experiments to astrophysical conditions, it is important to consider if the experiments give rise to multiphoton processes which are unlikely to occur in the ISM. 

We first consider the excitations of the PHP mode. 
Under the assumption of rapid intermolecular energy transfer, we can expect full relaxation of an excited PHP molecule between the individual few ps long micropulses that are spaced $1$\,ns apart.
It is then necessary to consider the density of photons absorbed in each micropulse. The PHP irradiation for the layered system with irradiation energy of $61$\,mJ has the highest photon flux with $1.4\times 10^{20}$\,photons\,cm$^{-2}$ in each macropulse and thus $2.3\times 10^{16}$\,photons\,cm$^{-2}$ in each micropulse. With an estimated PHP column density of $N_\mathrm{PHP}=1.8\times 10^{15}$\,molecules\,cm$^{-2}$, there are ${\sim}\,12.7$ incident photons per PHP molecule. We estimate the fraction of incident photons absorbed by the ice as $\chi = 1-10^{-\mathrm{Abs}}$, obtained from the experimental absorbance at the relevant wavelength. This should only be considered an estimate given the difference between the angle of incidence for the RAIRS experiments using unpolarized light ($80\degree$) and the \textit{p}-polarized FEL beam ($45\degree$). We do not consider the width of the FEL beam in this estimate as the full beam width is covered by the PHP bands.
Based on the experimental spectrum, a fraction of $\chi=0.025$ photons are absorbed in this case. Therefore, an estimated 0.32 photons are absorbed per PHP molecule in each micropulse in the PHP irradiation of the layered ice. For the mixed ice, an equivalent analysis gives 0.24 absorbed photons per PHP molecule. This estimate shows that up to one in every three PHP molecules could be excited in each micropulse. So for PHP irradiations, while multi-excitations of the same PHP molecule are unlikely, they cannot be ruled out, and multiphoton processes caused by interactions between neighbouring excited molecules can feasibly occur, especially for the layered case. 

Next, we consider excitation of the CO mode. The CO irradiation for the mixed system with an irradiation energy of $38$\,mJ has the highest photon flux with $0.63\times 10^{20}$\,photons\,cm$^{-2}$ in each macropulse and thus $1.0\times 10^{16}$\,photons\,cm$^{-2}$ in each micropulse.
With an estimated CO column density of $N_\mathrm{CO}=19.3\times 10^{15}$\,molecules\,cm$^{-2}$, there are $\sim 0.54$ incident photons per CO molecule.
As a result of the sharpness of the CO peak, the full width of the FEL beam does not lie within the CO band. We again determine the fraction of absorbed photons solely based on the absorbance at the irradiation wavelength, which is therefore an upper limit. Based on the experimental absorbance spectrum, $\chi = 0.13$. Therefore, an upper limit of 0.07 photons are absorbed per CO molecule in each micropulse for this irradiation. For the layered ice, an equivalent analysis gives 0.10 absorbed photons per CO molecule. \citet{Ferrari2024} modelled vibrational excitations of CO and reported inefficient energy dissipation and long lifetimes for vibrationally excited CO, on the order of ${\sim}\,1$\,ns. Based on this, it is feasible that an excited CO molecule could stay excited between micropulses, making multiphoton excitation events between subsequent micropulses possible.
These considerations show that while multiphoton processes within a micropulse are less likely for CO with an upper limit of $10\%$ CO molecules excited, we cannot rule out multiphoton effects between micropulses.

These simple estimates for photons per molecule assume an even distribution of energy across the irradiation area and use the total absorption for each ice to estimate average photon densities. We note that the photon density is higher in the centre of the irradiation spot, and that more of the photon absorption occurs in the upper layers of the ice.
Nevertheless, the presented arguments support that multi-excitations of the same molecule within a micropulse are unlikely to play a significant role in the observed desorption for any of the irradiations in this work. However, we cannot rule out some multiphoton effects between neighbouring excited molecules or even between subsequent micropulses in the case of CO irradiations. 
Our results can therefore not be extrapolated directly to ISM conditions without further studies. 

\end{appendix}

\end{document}